\pdfoutput=1

\documentclass[12pt,a4paper]{article}

\usepackage{ifthen} 
\newboolean{pdflatex}
\setboolean{pdflatex}{true} 

\newboolean{articletitles}
\setboolean{articletitles}{true} 

\newboolean{uprightparticles}
\setboolean{uprightparticles}{false} 

\newboolean{inbibliography}
\setboolean{inbibliography}{false} 

\usepackage{microtype}
\usepackage{lineno}  
\usepackage{xspace} 
\usepackage{caption} 

\usepackage{graphicx}  
\usepackage{color}
\usepackage{colortbl}
\graphicspath{{./figs/}} 

\usepackage{amsmath} 
\usepackage{amssymb}
\usepackage{amsfonts}
\usepackage{upgreek} 

\usepackage{bigstrut}

\newcommand*\patchAmsMathEnvironmentForLineno[1]{%
\expandafter\let\csname old#1\expandafter\endcsname\csname #1\endcsname
\expandafter\let\csname oldend#1\expandafter\endcsname\csname
end#1\endcsname
 \renewenvironment{#1}%
   {\linenomath\csname old#1\endcsname}%
   {\csname oldend#1\endcsname\endlinenomath}%
}
\newcommand*\patchBothAmsMathEnvironmentsForLineno[1]{%
  \patchAmsMathEnvironmentForLineno{#1}%
  \patchAmsMathEnvironmentForLineno{#1*}%
}
\AtBeginDocument{%
\patchBothAmsMathEnvironmentsForLineno{equation}%
\patchBothAmsMathEnvironmentsForLineno{align}%
\patchBothAmsMathEnvironmentsForLineno{flalign}%
\patchBothAmsMathEnvironmentsForLineno{alignat}%
\patchBothAmsMathEnvironmentsForLineno{gather}%
\patchBothAmsMathEnvironmentsForLineno{multline}%
\patchBothAmsMathEnvironmentsForLineno{eqnarray}%
}

\usepackage{hyperref}    
\usepackage[all]{hypcap} 

\def\lhcb {\mbox{LHCb}\xspace}

\def\cms    {\mbox{CMS}\xspace}

\def\dzero  {\mbox{D0}\xspace}

\def\MagUp {\mbox{\em Mag\kern -0.05em Up}\xspace}

\ifthenelse{\boolean{uprightparticles}}%
{

 \def\Pmu         {\ensuremath{\upmu}\xspace}

 \def\Ppi         {\ensuremath{\uppi}\xspace}

 \def\Ppsi        {\ensuremath{\uppsi}\xspace}

 \def\PDelta      {\ensuremath{\Delta}\xspace}                 
 \def\PXi      {\ensuremath{\Xi}\xspace}                 
 \def\PLambda      {\ensuremath{\Lambda}\xspace}                 
 \def\PSigma      {\ensuremath{\Sigma}\xspace}                 
 \def\POmega      {\ensuremath{\Omega}\xspace}                 
 \def\PUpsilon      {\ensuremath{\Upsilon}\xspace}                 
 
 \def\PB      {\ensuremath{\mathrm{B}}\xspace}                 
                  
 \def\PD      {\ensuremath{\mathrm{D}}\xspace}

 \def\PJ      {\ensuremath{\mathrm{J}}\xspace}                 
 \def\PK      {\ensuremath{\mathrm{K}}\xspace}

 \def\Pb      {\ensuremath{\mathrm{b}}\xspace}                 
 \def\Pc      {\ensuremath{\mathrm{c}}\xspace}                 
                  
 \def\Pe      {\ensuremath{\mathrm{e}}\xspace}

 \def\Pi      {\ensuremath{\mathrm{i}}\xspace}

 \def\Pp      {\ensuremath{\mathrm{p}}\xspace}

 \def\Ps      {\ensuremath{\mathrm{s}}\xspace}                 
                  
 \def\Pu      {\ensuremath{\mathrm{u}}\xspace}

}
{

 \def\Pmu         {\ensuremath{\mu}\xspace}

 \def\Ppi         {\ensuremath{\pi}\xspace}

 \def\Ppsi        {\ensuremath{\psi}\xspace}                 
                  
 \mathchardef\PDelta="7101
 \mathchardef\PXi="7104
 \mathchardef\PLambda="7103
 \mathchardef\PSigma="7106
 \mathchardef\POmega="710A
 \mathchardef\PUpsilon="7107
                  
 \def\PB      {\ensuremath{B}\xspace}                 
                  
 \def\PD      {\ensuremath{D}\xspace}

 \def\PJ      {\ensuremath{J}\xspace}                 
 \def\PK      {\ensuremath{K}\xspace}

 \def\Pb      {\ensuremath{b}\xspace}                 
 \def\Pc      {\ensuremath{c}\xspace}                 
                  
 \def\Pe      {\ensuremath{e}\xspace}

 \def\Pi      {\ensuremath{i}\xspace}

 \def\Pp      {\ensuremath{p}\xspace}

 \def\Ps      {\ensuremath{s}\xspace}                 
                  
 \def\Pu      {\ensuremath{u}\xspace}

}

\makeatletter
  \newcommand{\miniscule}{\@setfontsize\miniscule{4}{5}}
\makeatother

\DeclareRobustCommand{\optbar}[1]{\shortstack{{\miniscule (\rule[.5ex]{1.25em}{.18mm})}
  \\ [-.7ex] $#1$}}

\def\epem       {{\ensuremath{\Pe^+\Pe^-}}\xspace}

\def\mumu       {{\ensuremath{\Pmu^+\Pmu^-}}\xspace}

\def\ellm       {{\ensuremath{\ell^-}}\xspace}
\def\ellp       {{\ensuremath{\ell^+}}\xspace}

\def\uquark    {{\ensuremath{\Pu}}\xspace}

\def\squark    {{\ensuremath{\Ps}}\xspace}
\def\squarkbar {{\ensuremath{\overline \squark}}\xspace}

\def\cquark    {{\ensuremath{\Pc}}\xspace}
\def\cquarkbar {{\ensuremath{\overline \cquark}}\xspace}

\def\bquark    {{\ensuremath{\Pb}}\xspace}
\def\bquarkbar {{\ensuremath{\overline \bquark}}\xspace}

\def\pion   {{\ensuremath{\Ppi}}\xspace}

\def\pip    {{\ensuremath{\pion^+}}\xspace}
\def\pim    {{\ensuremath{\pion^-}}\xspace}
\def\pipm   {{\ensuremath{\pion^\pm}}\xspace}

\def\kaon    {{\ensuremath{\PK}}\xspace}
  \def\Kbar    {{\kern 0.2em\overline{\kern -0.2em \PK}{}}\xspace}

\def\KorKbar    {\kern 0.18em\optbar{\kern -0.18em K}{}\xspace}
\def\Kz      {{\ensuremath{\kaon^0}}\xspace}

\def\Kp      {{\ensuremath{\kaon^+}}\xspace}
\def\Km      {{\ensuremath{\kaon^-}}\xspace}
\def\Kpm     {{\ensuremath{\kaon^\pm}}\xspace}

\def\KS      {{\ensuremath{\kaon^0_{\rm\scriptscriptstyle S}}}\xspace}

\def\Kstarz  {{\ensuremath{\kaon^{*0}}}\xspace}
\def\Kstarzb {{\ensuremath{\Kbar{}^{*0}}}\xspace}
\def\Kstar   {{\ensuremath{\kaon^*}}\xspace}

\def\Dbar    {{\kern 0.2em\overline{\kern -0.2em \PD}{}}\xspace}
\def\D       {{\ensuremath{\PD}}\xspace}

\def\DorDbar    {\kern 0.18em\optbar{\kern -0.18em D}{}\xspace}
\def\Dz      {{\ensuremath{\D^0}}\xspace}
\def\Dzb     {{\ensuremath{\Dbar{}^0}}\xspace}

\def\Dsp     {{\ensuremath{\D^+_\squark}}\xspace}
\def\Dsm     {{\ensuremath{\D^-_\squark}}\xspace}
\def\Dspm    {{\ensuremath{\D^{\pm}_\squark}}\xspace}

\def\B       {{\ensuremath{\PB}}\xspace}
\def\Bbar    {{\ensuremath{\kern 0.18em\overline{\kern -0.18em \PB}{}}}\xspace}
\def\Bb      {{\ensuremath{\Bbar}}\xspace}
\def\BorBbar    {\kern 0.18em\optbar{\kern -0.18em B}{}\xspace}
\def\Bz      {{\ensuremath{\B^0}}\xspace}

\def\Bu      {{\ensuremath{\B^+}}\xspace}

\def\Bp      {{\ensuremath{\Bu}}\xspace}

\def\Bpm     {{\ensuremath{\B^\pm}}\xspace}

\def\Bd      {{\ensuremath{\B^0}}\xspace}
\def\Bs      {{\ensuremath{\B^0_\squark}}\xspace}

\def\jpsi     {{\ensuremath{{\PJ\mskip -3mu/\mskip -2mu\Ppsi\mskip 2mu}}}\xspace}

  \def\Y#1S{\ensuremath{\PUpsilon{(#1S)}}\xspace}

\def\proton      {{\ensuremath{\Pp}}\xspace}
\def\antiproton  {{\ensuremath{\overline \proton}}\xspace}

\def\Lz          {{\ensuremath{\PLambda}}\xspace}
\def\Lbar        {{\ensuremath{\kern 0.1em\overline{\kern -0.1em\PLambda}}}\xspace}
\def\LorLbar    {\kern 0.18em\optbar{\kern -0.18em \PLambda}{}\xspace}

\def\Lb      {{\ensuremath{\Lz^0_\bquark}}\xspace}

\newcommand{\decay}[2]{\ensuremath{#1\!\to #2}\xspace}         

\def\to                 {\ensuremath{\rightarrow}\xspace}

\def\CP                {{\ensuremath{C\!P}}\xspace}
\def\CPT               {{\ensuremath{C\!PT}}\xspace}

\def\AT#1     {\ensuremath{A_{\mathrm{T}}^{#1}}\xspace}           

\def\C#1      {\ensuremath{\mathcal{C}_{#1}}\xspace}                       
\def\Cp#1     {\ensuremath{\mathcal{C}_{#1}^{'}}\xspace}                    
\def\Ceff#1   {\ensuremath{\mathcal{C}_{#1}^{\mathrm{(eff)}}}\xspace}        
\def\Cpeff#1  {\ensuremath{\mathcal{C}_{#1}^{'\mathrm{(eff)}}}\xspace}       
\def\Ope#1    {\ensuremath{\mathcal{O}_{#1}}\xspace}                       
\def\Opep#1   {\ensuremath{\mathcal{O}_{#1}^{'}}\xspace}                    

\newcommand{\tev}{\ifthenelse{\boolean{inbibliography}}{\ensuremath{~T\kern -0.05em eV}\xspace}{\ensuremath{\mathrm{\,Te\kern -0.1em V}}}\xspace}
\newcommand{\gev}{\ensuremath{\mathrm{\,Ge\kern -0.1em V}}\xspace}
\newcommand{\mev}{\ensuremath{\mathrm{\,Me\kern -0.1em V}}\xspace}
\newcommand{\kev}{\ensuremath{\mathrm{\,ke\kern -0.1em V}}\xspace}
\newcommand{\ev}{\ensuremath{\mathrm{\,e\kern -0.1em V}}\xspace}
\newcommand{\gevc}{\ensuremath{{\mathrm{\,Ge\kern -0.1em V\!/}c}}\xspace}
\newcommand{\mevc}{\ensuremath{{\mathrm{\,Me\kern -0.1em V\!/}c}}\xspace}
\newcommand{\gevcc}{\ensuremath{{\mathrm{\,Ge\kern -0.1em V\!/}c^2}}\xspace}
\newcommand{\gevgevcccc}{\ensuremath{{\mathrm{\,Ge\kern -0.1em V^2\!/}c^4}}\xspace}
\newcommand{\mevcc}{\ensuremath{{\mathrm{\,Me\kern -0.1em V\!/}c^2}}\xspace}

\def\cm   {\ensuremath{\rm \,cm}\xspace}

\def\mm   {\ensuremath{\rm \,mm}\xspace}

\def\fb   {\ensuremath{\mbox{\,fb}}\xspace}
\def\invfb   {\ensuremath{\mbox{\,fb}^{-1}}\xspace}

\def\invps{\ensuremath{{\rm \,ps^{-1}}}\xspace}

\def\gsim{{~\raise.15em\hbox{$>$}\kern-.85em
          \lower.35em\hbox{$\sim$}~}\xspace}
\def\lsim{{~\raise.15em\hbox{$<$}\kern-.85em
          \lower.35em\hbox{$\sim$}~}\xspace}

\def\degrees{\ensuremath{^{\circ}}\xspace}

\def\rad{\ensuremath{\rm \,rad}\xspace}

\def\tell1  {TELL1\xspace}
\def\ukl1   {UKL1\xspace}

\usepackage{cite} 
\usepackage{mciteplus}

\newcommand{\aerr}[2]{{\:}^{+{\:}#1}_{-{\:}#2}}%

\def\lhcb{LHCb\xspace}
\def\cms{CMS\xspace}
\def\bds{\ensuremath{B_{(s)}^0}\xspace}
\def\bd{\ensuremath{B^0}\xspace}
\def\bs{\ensuremath{B_{s}^0}\xspace}

\def\bdsmumu{\ensuremath{B_{(s)}^0 \rightarrow \mu^+\mu^-}\xspace}
\def\bdmumu{\ensuremath{B^0 \rightarrow \mu^+\mu^-}\xspace}
\def\bsmumu{\ensuremath{B_{s}^0 \rightarrow \mu^+\mu^-}\xspace}

\def\fb{\ensuremath{\,\mathrm{fb}^{-1}}\xspace}

\def\Bhhprime{\ensuremath{B^0_{(s)}\to h^+h^{'-}}\xspace}

\def\bdkpi{\ensuremath{B^0\to K^+ \pi^-}\xspace}
\def\bujpsik{\ensuremath{B^+\to J/\psi K^+}\xspace}

\def\Bhmumu{\ensuremath{B \to h \mu\mu}\xspace}
\def\Bhmunu{\ensuremath{B \to h \mu\nu}\xspace}
\def\Lbpmunu{\ensuremath{\Lambda^0_b \to p \mu^-\bar{\nu}}\xspace}

\def\fb{\ensuremath{\,\mathrm{fb}^{-1}}\xspace}

\def\mmumu{\ensuremath{m_{\mu\mu}}\xspace}

\def\fsfd{\ensuremath{f_s/f_d}\xspace}

\def\br[#1]#2{\ensuremath{\mathcal{B}(#1) = #2}\xspace}
\def\R[#1]#2{\ensuremath{\mathcal{R}^{#1} = #2}\xspace}
\newcommand{\BRof}[1]{\ensuremath{{\cal B}(#1)}\xspace}

\def\RB{\ensuremath{{\mathcal{R}}}\xspace}

\newcommand{\RBSM}{\ensuremath{0.0295^{\:+\:0.0028}_{\:-\:0.0025}}\xspace}
\def\bsmumumeas{\ensuremath{\left( 2.8 ^{\:+\:0.7}_{\:-\:0.6} \right) \times 10^{-9}}\xspace}
\def\bdmumumeas{\ensuremath{\left( 3.9 ^{\:+\:1.6}_{\:-\:1.4} \right) \times 10^{-10}}\xspace}

\usepackage{longtable} 

\begin{document}

\renewcommand{\thefootnote}{\fnsymbol{footnote}}
\setcounter{footnote}{1}

\begin{titlepage}
\pagenumbering{roman}

\noindent

\vspace*{2.0cm}

{\normalfont\bfseries\boldmath\huge
\begin{center}
  Precision physics with heavy-flavoured hadrons
\end{center}
}

\vspace*{1.0cm}

\begin{center}
\large Patrick Koppenburg\\
\bigskip
{\it
\footnotesize
Nikhef National Institute for Subatomic Physics\\
PO box 41882, 1009 DB Amsterdam,The Netherlands\\
\smallskip
\smallskip
European Organization for Nuclear Research (CERN)\\
CH-1211 Geneva 23, Switzerland\\
\rm Patrick.Koppenburg@cern.ch}\\
\bigskip
\bigskip
\large Vincenzo Vagnoni\\
\bigskip
{\it
\footnotesize
Istituto Nazionale di Fisica Nucleare (INFN), Sezione di Bologna\\
via Irnerio 46, 40126 Bologna, Italy\\
\rm Vincenzo.Vagnoni@bo.infn.it}
\end{center}
\vspace*{1cm}

\begin{abstract}
  \noindent
 The understanding of flavour dynamics is one of the key aims of elementary particle physics. The last 15 years have witnessed the triumph of the Kobayashi-Maskawa mechanism, which describes all flavour changing transitions of quarks in the Standard Model. This important milestone has been reached owing to a series of experiments, in particular to those operating at the so-called $B$ factories, at the Tevatron, and now at the LHC. We briefly review status and perspectives of flavour physics, highlighting the results where the LHC has given the most significant contributions, notably including the recent observation of the $B_s^0\to\mu^+\mu^-$ decay. 
\end{abstract}

\vspace*{1.0cm}

\begin{center}
  Published in {\it 60 Years of CERN Experiments and Discoveries}\\
  \smallskip
  \smallskip
  {\small Advanced Series on Directions in High Energy Physics, Vol. 23, World Scientific Publishing}
\end{center}

\vspace{\fill}

\vspace*{2mm}

\end{titlepage}


\newpage
\setcounter{page}{2}
\mbox{~}

\cleardoublepage

\renewcommand{\thefootnote}{\arabic{footnote}}
\setcounter{footnote}{0}

\pagestyle{plain} 
\setcounter{page}{1}
\pagenumbering{arabic}

\section{Introduction}\label{sec:intro}
Flavour physics has played a central r\^ole in the development of the Standard Model (SM), which represents the state of the art of the fundamental theory of elementary physics interactions. The SM is able to describe with excellent accuracy all of the fundamental physics phenomena related to the electromagnetic, weak and strong forces, observed to date. Yet, it fails with some key aspects, notably including the fact that it does not provide an answer to one of the most fundamental questions: why is antimatter absent from the observed universe? Owing to the work of Andrei Sakharov in 1967~\cite{Sakharov:1967dj}, the phenomenon of $C\!P$ violation, \emph{i.e.} the non-invariance of the laws of nature under the combined application of charge ($C$) and parity ($P$) transformations, is known to be one of the ingredients needed to dynamically generate a baryon asymmetry starting from an initially symmetric universe.
However, it is also known that the size of \CP violation in the SM is too small, by several orders of magnitude, to explain the observed baryon asymmetry of the universe~\cite{Cohen:1993nk,Riotto:1999yt,Hou:2008xd}. 
As a consequence, other sources of \CP violation beyond the SM (BSM), which should produce observable effects in the form of deviations from the SM predictions of certain \CP-violating quantities, must exist. Rare decays that are strongly suppressed in the SM are of particular interest, since BSM amplitudes could be relatively sizable with respect to those of the SM.

The first run of the Large Hadron Collider (LHC), with 7 and 8\:\tev $pp$ collisions, has led to the discovery of the Higgs boson~\cite{Aad:2012tfa,Chatrchyan:2012ufa}, but no hint of the existence of other new particles has been found. Neither supersymmetry nor any other direct sign of BSM physics has popped out of the data. Besides the Higgs discovery, analyses from the first years of running have also firmly established the great impact of the ATLAS~\cite{Aad:2008zzm}, CMS~\cite{Chatrchyan:2008aa} and LHCb~\cite{Alves:2008zz} experiments in the field of \CP violation and rare decays of heavy-flavoured hadrons. In particular, LHCb has produced a plethora of results on a broad range of flavour observables in the $c$- and $b$-quark sectors, and ATLAS and CMS have given significant contributions to the $b$-quark sector, mainly using final states containing muon pairs. Also these measurements do not provide hints of BSM physics.

Nevertheless, it is of fundamental importance for future developments of elementary particle physics to keep improving the theoretical and experimental fields of flavour physics. On the one hand, such improvements increase the reach of indirect searches for BSM physics, probing higher and higher mass scales in the event that no BSM effects were discovered by direct detection. On the other hand, they would enable the BSM Lagrangian to be precisely determined, if any new particle were detected in direct searches. Starting with a brief historical perspective on the development of heavy flavour physics, we review the present status, highlighting some of the results where the LHC has given the most significant contributions.

\section{An historical perspective}\label{sec:history}
\subsection{The origin of the KM mechanism}\label{sec:ckmhistory}
As already mentioned, flavour physics has played a prominent  r\^ole in the development of the SM. As an example, one of the most notable predictions made in this context was that of the existence of a third quark generation, in a famous paper of 1973 by Makoto Kobayashi and Toshihide Maskawa~\cite{Kobayashi:1973fv}. In that work, which won them the Nobel Price in Physics in 2008 ``for the discovery of the origin of the broken symmetry which predicts the existence of at least three families of quarks in nature'', Kobayashi and Maskawa extended the Cabibbo~\cite{Cabibbo:1963yz} (with only $u$, $d$, and $s$ quarks) and the Glashow-Iliopoulos-Maiani~\cite{Glashow:1970gm} (GIM, including also the $c$ quark) mechanisms, pointing out that $C\!P$ violation could be incorporated into the emerging picture of the SM if six quarks were present. This is commonly referred to as the Kobayashi-Maskawa (KM) mechanism. It must be emphasised that at that time only hadrons made of the three lighter quarks had been observed. An experimental revolution took place in 1974, when a new state containing the $c$ quark was discovered almost simultaneously at Brookhaven~\cite{Aubert:1974js} and SLAC~\cite{Augustin:1974xw}. Then, the experimental observations of the $b$~\cite{Herb:1977ek} and $t$~\cite{Abe:1995hr} quarks were made at FNAL in 1977 and 1995, respectively. 

The idea of Kobayashi and Maskawa, formalised in the so-called Cabibbo-Kobayashi-Maskawa (CKM) quark mixing matrix, was then included in the SM by the beginning of the 1980's. The phenomenon of $C\!P$ violation, first revealed in 1964 using decays of neutral kaons~\cite{Christenson:1964fg}, was elegantly accounted for as an irreducible complex phase within the CKM matrix. The experimental proof of the validity of the KM mechanism and the precise measurement of the value of the \CP-violating phase soon became questions of paramount importance.

\subsection{The rise of $B$ physics}\label{sec:BPhysicsHistory}
Due to the nature of the CKM matrix, an accurate test of the KM mechanism required an extension of the physics programme to heavy-flavoured hadrons. Pioneering steps in the $b$-quark flavour sector were moved at the beginning of the 1980's by the CLEO experiment at CESR~\cite{Andrews:1982dp}. At the same time, Ikaros Bigi, Ashton Carter and Tony Sanda published papers exploring the possibility that large $C\!P$-violating effects could be present in the decay rates of $B^0$ mesons decaying to the $J/\psi K^0_S$ \CP eigenstate~\cite{Carter:1980tk,Bigi:1983cj}. In addition, they also pointed out that such a measurement could be interpreted in terms of the $C\!P$-violating phase without relevant theoretical uncertainties due to strong interaction effects. However, there were two formidable obstacles to overcome: first, an experimental observation required an enormous amount of $B^0$ mesons, well beyond what was conceivable to produce and collect at the time; second, a precise measurement of the decay time was required, together with the knowledge of the flavour of the $B^0$ meson at production.

Few years later, in 1987, the ARGUS experiment at DESY measured for the first time the mixing rate of $B^0$ and $\bar{B}^0$ mesons~\cite{Albrecht:1988vy}, whose knowledge was an important ingredient to understand the feasibility of measuring $C\!P$ violation with $B^0 \to J/\psi K^0_S$ decays. Another crucial ingredient came along due to tremendous developments in the performance of $e^+e^-$ storage rings. By the late 1980's, many different possible designs for new machines were being explored. A novel idea was put forward by Pier Oddone in 1987: a high-luminosity asymmetrical $e^+e^-$ circular collider operating at the centre-of-mass energy of the $\Upsilon(4S)$ meson~\cite{Oddone:1987up}. Owing to the beam-energy asymmetry, $B$ mesons would be produced with a boost in the laboratory frame towards the direction of the most energetic beam. The consequent nonzero decay length, measured by means of state-of-the-art silicon vertex detectors, would enable precise measurements of the decay time to be achieved. Two machines based on Oddone's concept, so-called $B$ factories, were eventually built: PEP-II at SLAC in the United States and KEKB at KEK in Japan. The associated detectors, BaBar~\cite{Aubert:2001tu} at PEP-II and Belle~\cite{Abashian:2000cg} at KEKB, were approved in 1993 and in 1994, respectively. If CESR was initially able to produce few tens of $b\bar{b}$ pairs per day, PEP-II and KEKB were capable of producing order of one million $b\bar{b}$ pairs per day.

Meanwhile, during the course of the 1990's, many $b$-physics measurements were being performed at the $Z^0$ factories, \emph{i.e.} the LEP experiments~\cite{Decamp:1990jra,Aarnio:1990vx,L3:1989aa,Ahmet:1990eg} at CERN, and the SLD~\cite{SLD:1984aa} experiment at SLAC. Despite the relatively small statistics, $b$ hadrons produced in $Z^0$ decays were naturally characterised by a large boost, enabling measurements of lifetimes of all $b$-hadron species and of oscillation frequencies of neutral $B$ mesons to be performed. In particular, for the first time it was possible to study samples of $B^0_s$-meson, $b$-baryon and even a handful of $B^+_c$-meson decays~\cite{Barker:2010iva,Rowson:2001cd}. Similar pioneering measurements were also made at the Tevatron with Run I data, using hadronic collisions as a source of $b$ quarks~\cite{Paulini:1999px}.

Soon after PEP-II and KEKB were turned on, the two machines broke any existing record of instantaneous and integrated luminosity of previous particle colliders. By the end of their research programmes, BaBar and Belle measured $C\!P$ violation in $B^0 \to J/\psi K^0_S$ decays with a relative precision of about 3\%~\cite{Aubert:2009aw,Adachi:2012et}.
The large sample of $B$-meson decays collected at BaBar and Belle enabled a series of further measurements in the flavour sector to be performed, well beyond the initial expectations. In the same years, a major step forward in these topics was also made at the Tevatron with Run II data. Although with a somewhat limited scope if compared to $B$ factories, the CDF and D0 experiments at FNAL collected large amounts of heavy-flavoured-hadron decays, performing some high precision measurements~\cite{Kuhr:2013hd}, notably including the first observation of $B^0_s$-meson mixing in 2006~\cite{Abulencia:2006ze}. 

\subsection{The LHC era}\label{sec:LHCb}
When the constructions of the BaBar and Belle detectors were being scrutinised for approval, three distinct proposals for a dedicated $b$-physics experiment at the LHC were put forward, so-called COBEX, GAJET, and LHB. GAJET and LHB were both based on fixed targets, the former working with a gas target placed inside the LHC beam pipe and the latter exploiting an extracted LHC beam. COBEX was instead proposed to work in proton-proton collider mode. The three groups of proponents were asked to join together and submit to the LHC Experiments Committee (LHCC) a proposal for a single collider-mode experiment, namely LHCb~\cite{Alves:2008zz}. LHCb was then designed to exploit the potential for heavy-flavour physics at the LHC by instrumenting the forward region of proton-proton collisions, in order to take advantage of the large $b\bar{b}$ cross section in the forward (or backward) LHC beam direction. The LHCb experiment was approved in 1998, and started taking data with the start-up of LHC in 2009. 

\begin{figure}[tb]
  \begin{center}
    \includegraphics[width=0.9\textwidth]{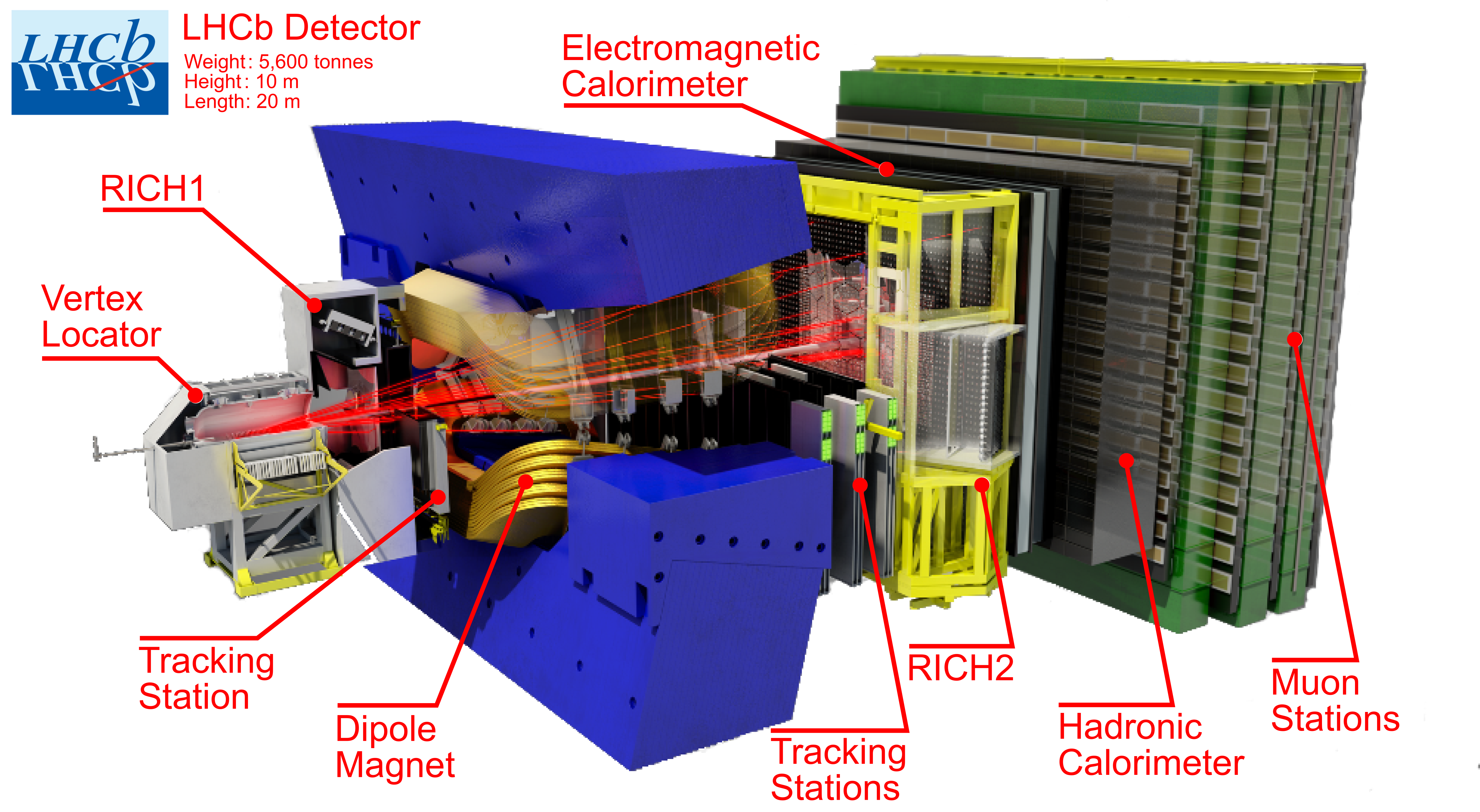}
  \end{center}
  \caption{Sketch of the LHCb detector.}\label{Fig:LHCb}
\end{figure}
The LHCb detector~\cite{Alves:2008zz,LHCb-DP-2014-002}, shown in Fig.~\ref{Fig:LHCb},
includes a high-precision tracking system
consisting of a silicon-strip vertex detector surrounding the $pp$
interaction region, a large-area silicon-strip detector located
upstream of a dipole magnet with a bending power of about
$4{\rm\,Tm}$, and three stations of silicon-strip detectors and straw
drift tubes placed downstream of the magnet.
The tracking system provides a measurement of momentum of charged particles with
a relative uncertainty that varies from 0.5\% at low momentum to 1.0\% at 200\gevc.
The silicon sensors of the vertex detector come
as close as 8\:\mm to the LHC beam. This allows for a very precise measurement
of the track trajectory close to the interaction point, which is crucial to 
separate decays of beauty and charm hadrons, with typical 
flight distances of a few millimetres in the laboratory frame, from the background.
The distance of a track to a primary vertex, the impact parameter,
is measured with a resolution of 15--30\:$\mu$m.

One distinctive feature of LHCb, when compared to the ATLAS and CMS detectors, is its
particle identification capability for charged hadrons. This is mainly achieved by means of two ring-imaging Cherenkov (``RICH'') detectors
placed on either side of the tracking stations. Once particle momenta are measured, the two RICH detectors enable the identification of protons, kaons and pions to be obtained. An electromagnetic calorimeter, complemented with scintillating-pad and preshower detectors, provides energy and position of photons and electrons, and allow for their identification in conjunction with information from the tracking system. The electromagnetic calorimeter is followed by a hadronic calorimeter that also gives some information to identify hadrons.
Finally, muons are identified by a system composed of alternating layers of iron and multiwire proportional chambers.
The online event selection is performed by a trigger~\cite{LHCb-DP-2012-004} which consists of a hardware stage, based on information from the calorimeter and muon systems, followed by a software stage, which applies full event reconstruction. 

LHCb operates at a much lower instantaneous luminosity than the peak luminosity made available by the LHC. This is necessary
to better distinguish charged particles resulting from \bquark- and \cquark-hadron decays from
particles produced in other $pp$ collisions, in the forward region covered by the detector acceptance.
During the first run of the LHC, the average number of $pp$ collisions per bunch crossing at the LHCb intersection
was kept at about $1.7$, corresponding to a luminosity of $4\times10^{32}\rm \cm^{-2}s^{-1}$. This was achieved by a dynamical adjustment of the transverse offset between the LHC beams during the fill, enabling constant luminosity to be kept throughout.

Besides LHCb, the general purpose ATLAS and CMS detectors were also designed with the aim of performing $b$-physics measurements, mainly using final states containing muon pairs due to constraints dictated by the trigger and to the absence of sub-detectors with strong particle identification capabilities for charged hadrons.

\section{The CKM matrix}\label{sec:CKM}
\subsection{Definition}\label{sec:CKMdefinition}
In the SM, charged-current interactions of quarks are described by the Lagrangian
\begin{equation*}\label{eq:chargeInteraction}
  \mathcal{L}_{W^{\pm}} = - \frac{g}{\sqrt{2}}\overline{U}_{i}\gamma^{\mu}\frac{1-\gamma^5}{2}\left(V_{\rm CKM} \right )_{ij} D_{j} W_{\mu}^{+} + h.c.,
\end{equation*}
where $g$ is the electroweak coupling constant and $V_{\rm CKM}$ is the CKM matrix
\begin{equation*}\label{eq:ckmMatrix}
  V_{\rm CKM} = \left( 
    \begin{array}{lcr}
      V_{ud} & V_{us} & V_{ub} \\
      V_{cd} & V_{cs} & V_{cb} \\
      V_{td} & V_{ts} & V_{tb}
    \end{array}
  \right),
\end{equation*}
originating from the misalignment in flavour space of the up and down components of the $SU(2)_L$ quark doublet of the SM. The $V_{ij}$ matrix elements represent the couplings between up-type quarks $U_i=(u,\,c,\,t)$ and down-type quarks $D_j=(d,\,s,\,b)$.

An important feature of the CKM matrix is its unitarity. Such a condition determines the number of free parameters of the matrix. A generic $N\times N$ unitary matrix depends on $N\left(N-1\right)/2$ mixing angles and $N\left(N+1\right)/2$ complex phases. In the CKM case, dealing with a mixing matrix between the quark flavour eigenstates, the Lagrangian enables the phase of each quark field to be redefined, such that $2N-1$ unphysical phases cancel out. As a consequence, any $N\times N$ complex matrix describing mixing between $N$ generations of quarks has
\begin{equation*}\label{eq:freeParameter}
  \underbrace{\frac{1}{2}N\left(N-1\right)}_{\mbox{mixing angles}} \,\,\,\,\,\, + \underbrace{\frac{1}{2}\left(N-1\right)\left(N-2\right)}_{\mbox{physical complex phases}} = \,\,\,\,\,\, \left(N-1\right)^{2}
\end{equation*}
free parameters. The interesting case $N=2$ leads to the GIM mixing matrix with only one free parameter, namely the Cabibbo angle $\theta_{C}$~\cite{Cabibbo:1963yz}
\begin{equation*}\label{eq:cabibboMatrix}
  V_{\rm GIM}=\left(
    \begin{array}{lcr}
      \phantom{-}\cos{\theta_C} & \phantom{-}\sin{\theta_C} \\
      -\sin{\theta_C} & \phantom{-}\cos{\theta_C}
    \end{array}
  \right).
\end{equation*}
When formalised in 1970, the nature of $V_{\rm GIM}$ was invoked to explain the suppression of flavour changing neutral current (FCNC) processes, and put the basis for the discovery of the charm quark~\cite{Glashow:1970gm,Aubert:1974js,Augustin:1974xw}. In the case $N=3$, the resulting number of free parameters is four: three mixing angles and one complex phase. This phase alone is responsible for \CP violation in the weak interactions of the SM.

\subsection{Standard parametrisation}\label{sec:CKMStandard}

Among the various possible conventions, a standard choice to parametrise $V_{\rm CKM}$ is given by
\begin{equation*}\label{eq:ckmConvention}
  V_{\rm CKM} = \left(
    \begin{array}{ccc}
      c_{12}c_{13}                                          & s_{12}c_{13}                                          & s_{13}e^{-i\delta} \\
      -s_{12}c_{23}-c_{12}s_{23}s_{13}e^{i\delta} & c_{12}c_{23}-s_{12}s_{23}s_{13}e^{i\delta}   & s_{23}c_{13} \\
      s_{12}s_{23}-c_{12}c_{23}s_{13}e^{i\delta}   & -c_{12}s_{23}-s_{12}c_{23}s_{13}e^{i\delta} & c_{23}c_{13}
    \end{array}
  \right),
\end{equation*}
where $s_{ij} = \sin{\theta_{ij}}$, $c_{ij}=\cos{\theta_{ij}}$ and $\delta$ is the \CP-violating phase. All the $\theta_{ij}$ angles can be chosen to lie in the first quadrant, thus $s_{ij},c_{ij}\geq 0$. The coupling between quark generations $i$ and $j$ vanishes if the corresponding $\theta_{ij}$ is equal to zero. In the case where $\theta_{13}=\theta_{23}=0$, the third generation would decouple and the CKM matrix would take the form of $V_{\rm GIM}$. The presence of a complex phase in the mixing matrix is a necessary condition for \CP violation, although not sufficient. As pointed out in Ref.~\cite{Jarlskog:1985ht}, another key condition is
\begin{equation*}\label{eq:jarlskogCondition} \left(m_{t}^{2}-m_{c}^{2}\right)\left(m_{t}^{2}-m_{u}^{2}\right)\left(m_{c}^{2}-m_{u}^{2}\right)\left(m_{b}^{2}-m_{s}^{2}\right)\left(m_{b}^{2}-m_{d}^{2}\right)\left(m_{s}^{2}-m_{d}^{2}\right)\times J_{C\!P} \neq 0,
\end{equation*} 
where
\begin{equation*}\label{eq:jarlskogArea}
J_{C\!P} = \left|\Im\left(V_{i\alpha} V_{j\beta} V_{i\beta}^{*}V_{j\alpha}^{*}\right)\right|\qquad \left( i\neq j, \alpha\neq\beta\right)
\end{equation*}
is the so-called Jarlskog parameter. This condition is related to the fact that the CKM phase could be eliminated if any of two quarks with the same charge were degenerate in mass. As a consequence, the origin of \CP violation in the SM is deeply connected to the origin of the quark mass hierarchy and to the number of fermion generations.

The Jarlskog parameter can be interpreted as a measure of the size of \CP violation in the SM. Its value does not depend on the phase convention of the quark fields, and adopting the standard parametrisation it can be written as
\begin{equation*}\label{eq:jarlskogValue}
J_{C\!P} = s_{12}s_{13}s_{23}c_{12}c_{23}c_{13}^{2}\sin{\delta}.
\end{equation*}
Experimentally one has $J_{C\!P} = \mathcal{O}\left(10^{-5}\right)$, which quantifies how small \CP violation is in the SM.

\subsection{Wolfenstein parametrisation}\label{sec:CKMWolfenstein}

Experimental information leads to the evidence that transitions within the same generation are characterised by $V_{\rm CKM}$ elements of $\mathcal{O}(1)$. Instead, those between the first and second generations are suppressed by a factor $\mathcal{O}(10^{-1})$; those between the second and third generations by a factor $\mathcal{O}(10^{-2})$; and those between the first and third generations by a factor $\mathcal{O}(10^{-3})$. It can be stated that
\begin{equation*}\label{eq:sinHierarchy}
s_{12} \simeq 0.22 \gg s_{23} = \mathcal{O}(10^{-2}) \gg s_{13} = \mathcal{O}(10^{-3}).
\end{equation*}
It is useful to introduce a parametrisation of the CKM matrix, whose original formulation was due to Wolfenstein~\cite{Wolfenstein:1983yz},  defining
\begin{align*}
s_{12} &= \lambda =  \frac{\left|V_{us}\right|}{\sqrt{\left|V_{ud}\right|^{2}+\left|V_{us}\right|^{2}}},\\
s_{23} &= A\lambda^{2} = \lambda\left|\frac{V_{cb}}{V_{us}}\right|,\\
s_{13}e^{-i\delta} &= A\lambda^{3}\left(\rho -i\eta\right) = V_{ub} .
\end{align*}
The CKM matrix can be re-written as a power expansion of the parameter $\lambda$ (which corresponds to $\sin{\theta_{C}}$)
\begin{equation*}\label{eq:ckmWolfenstein4}
\resizebox{1\hsize}{!}{$
V_{\rm CKM} = \left(
  \begin{array}{ccc}
    1-\frac{1}{2}\lambda^{2}-\frac{1}{8}\lambda^{4}                                   & \lambda                  & A\lambda^{3}\left(\rho -i\eta\right) \\
    -\lambda +\frac{1}{2}A^{2}\lambda^{5}\left[1-2(\rho +i\eta)\right]        & 1-\frac{1}{2}\lambda^{2}-\frac{1}{8}\lambda^{4}(1+4A^{2})     & A\lambda^{2}                                     \\
    A\lambda^{3}\left[1-(\rho +i\eta)\left(1-\frac{1}{2}\lambda^{2}\right)\right]~ & ~-A\lambda^{2}+\frac{1}{2}A\lambda^{4}\left[1-2(\rho +i\eta)\right]~        & ~1-\frac{1}{2}A^{2}\lambda^{4}
  \end{array}
\right),
$}
\end{equation*}
which is valid up to $\mathcal{O}\left(\lambda^{6}\right)$. With this parametrisation, the CKM matrix is complex, and hence \CP violation is allowed for, if and only if $\eta$ differs from zero.
To lowest order the Jarlskog parameter becomes
\begin{equation*}\label{eq:jarlskogValueWolf}
J_{C\!P} = \lambda^{6}A^{2}\eta,
\end{equation*}
and, as expected, is directly related to the \CP-violating parameter $\eta$.

\subsection{The unitarity triangle}\label{sec:UT}

\begin{figure}[tb]
  \begin{center}
    \includegraphics[width=1\textwidth]{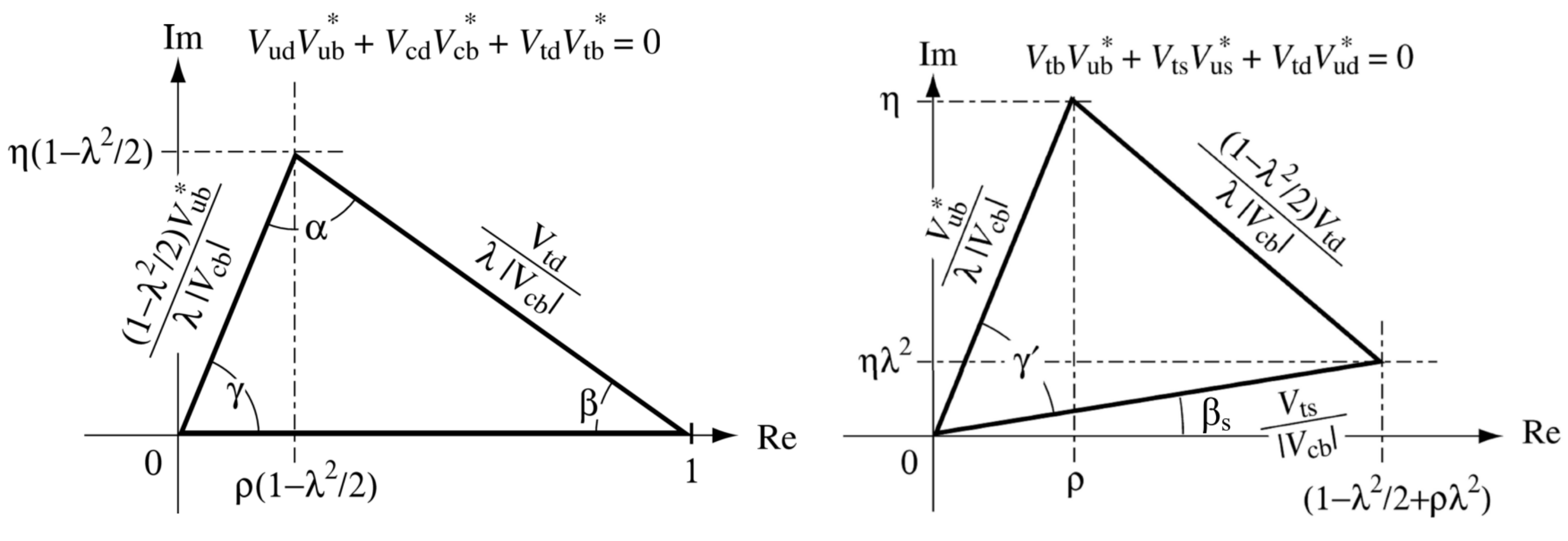}
  \end{center}
  \caption{Representation in the complex plane of the non-squashed triangles obtained from the off-diagonal unitarity relations of the CKM matrix.}\label{fig:unitaryTriangle}
\end{figure}

The unitarity condition of the CKM matrix, $V_{\rm CKM}V_{\rm CKM}^{\dagger} = V_{\rm CKM}^{\dagger}V_{\rm CKM} = \mathbb{I}$, leads to a set of 12 equations: 6 for diagonal terms and 6 for off-diagonal terms. In particular, the equations for the off-diagonal terms can be represented as triangles in the complex plane, all characterised by the same area $J_{C\!P}/2$
\begin{eqnarray*}
  \underbrace{V_{ud}V_{us}^{*}}_{\mathcal{O}(\lambda)}+\underbrace{V_{cd}V_{cs}^{*}}_{\mathcal{O}(\lambda)}+\underbrace{V_{td}V_{ts}^{*}}_{\mathcal{O}(\lambda^{5})} & = & 0, \\
  \underbrace{V_{us}V_{ub}^{*}}_{\mathcal{O}(\lambda^{4})}+\underbrace{V_{cs}V_{cb}^{*}}_{\mathcal{O}(\lambda^{2})}+\underbrace{V_{ts}V_{tb}^{*}}_{\mathcal{O}(\lambda^{2})} & = & 0, \\
  \underbrace{V_{ud}V_{ub}^{*}}_{\mathcal{O}(\lambda^{3})}+\underbrace{V_{cd}V_{cb}^{*}}_{\mathcal{O}(\lambda^{3})}+\underbrace{V_{td}V_{tb}^{*}}_{\mathcal{O}(\lambda^{3})} & = & 0, \\
  \underbrace{V_{ud}V_{cd}^{*}}_{\mathcal{O}(\lambda)}+\underbrace{V_{us}V_{cs}^{*}}_{\mathcal{O}(\lambda)}+\underbrace{V_{ub}V_{cb}^{*}}_{\mathcal{O}(\lambda^{5})} & = & 0, \\
  \underbrace{V_{cd}V_{td}^{*}}_{\mathcal{O}(\lambda^{4})}+\underbrace{V_{cs}V_{ts}^{*}}_{\mathcal{O}(\lambda^{2})}+\underbrace{V_{cb}V_{tb}^{*}}_{\mathcal{O}(\lambda^{2})} & = & 0, \\
  \underbrace{V_{ud}V_{td}^{*}}_{\mathcal{O}(\lambda^{3})}+\underbrace{V_{us}V_{ts}^{*}}_{\mathcal{O}(\lambda^{3})}+\underbrace{V_{ub}V_{tb}^{*}}_{\mathcal{O}(\lambda^{3})} & = & 0.
\end{eqnarray*}
Only two out of these six triangles have sides of the same order of magnitude, $\mathcal{O}(\lambda^{3})$. In terms of the Wolfenstein parametrisation, up to $\mathcal{O}(\lambda^{7})$ the corresponding equations can be written as
\begin{equation*}
    A\lambda^3\{(1-\lambda^2/2) (\rho+i\eta)+ [1-(1-\lambda^2/2) (\rho+i\eta) ] + (-1)\} = 0,
\end{equation*}
\begin{equation*}
  A\lambda^3\{ (\rho +i\eta) + [1-\rho -i\eta -\lambda^{2}(1/2-\rho -i\eta)] + [-1+\lambda^{2}(1/2 - \rho -i\eta)] \} = 0.
\end{equation*} 
Eliminating the common factor $A\lambda^{3}$ from both equations, the two triangles in the complex plane represented in Fig.~\ref{fig:unitaryTriangle} are obtained. In particular, the triangle defined by the former equation is commonly referred to as the unitarity triangle (UT).
The sides of the UT are given by
\begin{equation*}\label{eq:utRB}
  R_{u} \equiv \left|\frac{V_{ud}V_{ub}^{*}}{V_{cd}V_{cb}^{*}}\right|  =  \sqrt{\bar{\rho}^{2}+\bar{\eta}^{2}},
\end{equation*}
\begin{equation*}
  R_{t} \equiv \left|\frac{V_{td}V_{tb}^{*}}{V_{cd}V_{cb}^{*}}\right| = \sqrt{\left(1-\bar{\rho}\right)^{2}+\bar{\eta}^{2}}, \label{eq:utRT}
\end{equation*}
where to simplify the notation the parameters $\bar{\rho}$ and $\bar{\eta}$, namely the coordinates in the complex plane of the only non-trivial apex of the UT, the others being (0, 0) and (1, 0), have been introduced. The exact relation between $\bar{\rho}$ and $\bar{\eta}$ and the Wolfenstein parameters is defined by
\begin{equation*}
\rho + i\eta = \sqrt{\frac{1-A^2 \lambda^4}{1-\lambda^2}} \frac{\bar{\rho} + i\bar{\eta}}{1-A^2 \lambda^4(\bar{\rho}+i\bar{\eta})},
\end{equation*}
which, at the lowest non-trivial order in $\lambda$, yields
\begin{equation*}\label{eq:wolfensteinGeneralization}
\rho = \left(1+\frac{\lambda^{2}}{2}\right)\bar{\rho}+\mathcal{O}(\lambda^{4}),\qquad \eta = \left(1+\frac{\lambda^{2}}{2}\right)\bar{\eta}+\mathcal{O}(\lambda^{4}).
\end{equation*}
The angles of the UT are related to the CKM matrix elements as
\begin{equation*}
  \alpha\equiv \arg{\left(-\frac{V_{td}V_{tb}^{*}}{V_{ud}V_{ub}^{*}}\right)} = \arg{\left(-\frac{1-\bar{\rho}-i\bar{\eta}}{\bar{\rho}+i\bar{\eta}}\right)},\label{eq:alpha}
\end{equation*}
\begin{equation*}
  \beta \equiv\arg{\left(-\frac{V_{cd}V_{cb}^{*}}{V_{td}V_{tb}^{*}}\right)} = \arg{\left(\frac{1}{1-\bar{\rho}-i\bar{\eta}}\right)}, \label{eq:beta}
\end{equation*}
\begin{equation*}
  \gamma\equiv\arg{\left(-\frac{V_{ud}V_{ub}^{*}}{V_{cd}V_{cb}^{*}}\right)} = \arg{\left(\bar{\rho}+i\bar{\eta}\right)}. \label{eq:gamma}
\end{equation*}
The second non-squashed triangle has similar characteristics with respect to the UT. The apex is placed in the point $(\rho ,~\eta)$ and is tilted by an angle
\begin{equation*}\label{eq:betas}
  \beta_{s} \equiv \arg{\left(-\frac{V_{ts}V_{tb}^{*}}{V_{cs}V_{cb}^{*}}\right)} = \lambda^2 \eta + \mathcal{O}(\lambda^{4}).
\end{equation*}

\subsection{Phenomenology of \CP violation}

The phenomenon of \CP violation has been observed at a level above five standard deviations in a dozen of processes involving charged and neutral $B$-meson decays, as well as in a couple of neutral kaon decays~\cite{PDG2014}. In this section we focus in particular on the phenomenology of \CP violation in the $b$-quark sector.

Three types of \CP violation can occur in the quark sector: \CP violation in the decay (also known as direct \CP violation), \CP violation in the mixing of neutral mesons, and \CP violation in the interference between mixing and decay.

Defining the amplitude of a $B$ meson decaying to the final state $f$ as $A_f$, and that of its \CP conjugate $\bar{B}$ to the \CP conjugate final state $\bar{f}$ as $\bar{A}_{\bar{f}}$, direct \CP violation occurs when $\left |A_f \right| \neq \left| \bar{A}_{\bar{f}} \right |$. This is the only possible type of \CP violation that can be observed in the decays of charged mesons and baryons, where mixing is not allowed for.
If for example there are two distinct processes contributing to the decay amplitude, one can write
\begin{eqnarray*}\label{eq:amplitude}
  A_f & = & e^{i\varphi_{1}}\left|A_{1}\right|e^{i\delta_{1}} + e^{i\varphi_{2}}\left|A_{2}\right|e^{i\delta_{2}} \nonumber \\
  \bar{A}_{\bar{f}} & = & e^{-i\varphi_{1}}\left|A_{1}\right|e^{i\delta_{1}} + e^{-i\varphi_{2}}\left|A_{2}\right|e^{i\delta_{2}},
\end{eqnarray*}
where $\varphi_{1,2}$ denotes \CP-violating weak phases and $\left|A_{1,2}\right|e^{i\delta_{1,2}}$ \CP-conserving strong amplitudes of the two processes, labelled by the subscripts 1 and 2.
The \CP-violating asymmetry is given by
\begin{equation}\label{eq:directAsymmetry}
  \begin{split}
    A_{C\!P} & \equiv \frac{\Gamma_{\bar{B}\rightarrow\bar{f}} - \Gamma _{B\rightarrow f}}{\Gamma_{\bar{B}\rightarrow\bar{f}} +\Gamma _{B\rightarrow f}} =
    \frac{\left|\bar{A}_{\bar{f}}\right|^{2}-\left|A_{f}\right|^{2}}{\left|\bar{A}_{\bar{f}}\right|^{2}+\left|A_{f}\right|^{2}} = \\
    & = \frac{2\left|A_{1}\right|\left|A_{2}\right|\sin{\left(\delta_{2}-\delta_{1}\right)}\sin{\left(\varphi_{2}-\varphi_{1}\right)}}{\left|A_{1}\right|^{2}+2\left|A_{1}\right|\left|A_{2}\right|\cos{\left(\delta_{2}-\delta_{1}\right)}\cos{\left(\varphi_{2}-\varphi_{1}\right)}+\left|A_{2}\right|^{2}}.
  \end{split}
\end{equation}
A nonzero value of the asymmetry $A_{C\!P}$ arises from the interference between the two processes, and requires both a nonzero weak phase difference $\varphi_{2}-\varphi_{1}$ and a nonzero strong phase difference $\delta_{2}-\delta_{1}$. The presence of (at least) two interfering processes is a distinctive feature to observe \CP violation.

\begin{figure}[tb]
  \begin{center}
    \includegraphics[width=0.9\textwidth]{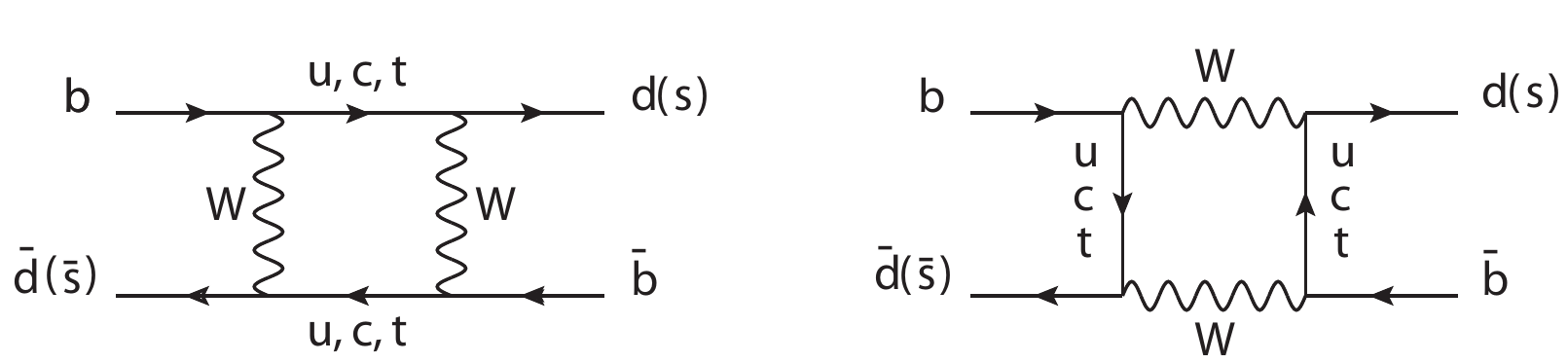}
  \end{center}
  \caption{Box diagrams contributing to $\bar{B}^{0}-B^{0}$ and $\bar{B}_{s}^{0}-B_{s}^{0}$ mixing.}\label{fig:boxDiagrams}
\end{figure}

When neutral heavy mesons are involved, the phenomenon of mixing between opposite flavours takes place. In the case of $B^0$ and $B^0_s$ mesons, due to the diagrams sketched in Fig.~\ref{fig:boxDiagrams}, an initial $\left| B \right\rangle$ state will evolve as a superposition of $\left| B \right\rangle$ and $\left| \overline{B} \right \rangle$ states.
Hence the mass eigenstates do not coincide with the flavour eigenstates, but are related to them by
\begin{equation*}\label{eq:massEigenstate}
    \left|B_\mathrm{H}\right\rangle = \frac{p\left|B\right\rangle +q\left|\bar{B}\right\rangle}{\sqrt{\left|p\right|^{2}+\left|q\right|^{2}}}, \qquad
    \left|B_\mathrm{L}\right\rangle = \frac{p\left|B\right\rangle -q\left|\bar{B}\right\rangle}{\sqrt{\left|p\right|^{2}+\left|q\right|^{2}}},
\end{equation*}
where $p$ and $q$ are two complex parameters, and $\left|B_\mathrm{H}\right\rangle$ and $\left|B_\mathrm{L}\right\rangle$ denote the two eigenstates of the $B^0_{(s)}$--$\Bb^0_{(s)}$  system. These two eigenstates are split in mass and lifetime, and we can define the mass and width differences $\Delta m_{d(s)} \equiv m_{{d(s)},\,\mathrm{H}}-m_{{d(s)},\,\mathrm{L}}$ and $\Delta\Gamma_{d(s)} \equiv \Gamma_{{d(s)},\,\mathrm{L}}-\Gamma_{{d(s)},\,\mathrm{H}}$. The subscripts $\mathrm{H}$ and $\mathrm{L}$ denote the heavy and light eigenstates. With this convention, the values of $\Delta m_{d}$ and $\Delta m_{s}$ are positive by definition. The present knowledge of $B^0$- and $B^0_s$-mixing processes is obtained from flavour-tagged time-dependent studies of semileptonic decays and of other decays involving flavour-specific final states, such as $B^0_s \to D_s^- \pi^+$. The world averages of the mass differences are $\Delta m_{d}=0.510 \pm 0.003\,\mathrm{ps}^{-1}$ and $\Delta m_{s}=17.757 \pm 0.021\,\mathrm{ps}^{-1}$~\cite{HFAG}. The value of $\Delta\Gamma_{s}$ is measured to be positive~\cite{Xie:2009fs,LHCb-PAPER-2011-028}, $\Delta\Gamma_s = 0.106 \pm 0.011\,(\mathrm{stat})\pm0.007\,(\mathrm{syst})\invps$~\cite{LHCb-PAPER-2011-028}. The value of $\Delta\Gamma_{d}$ is also positive in the SM and is expected to be much smaller than that of $\Delta\Gamma_{s}$, $\Delta\Gamma_{d} \simeq 3\times10^{-3}\invps$~\cite{Bona:2006ah}.

\CP violation in the mixing of neutral mesons arises when the rate of \emph{e.g.} $B^0_{(s)}$ mesons transforming into $\Bb^0_{(s)}$ mesons differs from the rate of $\Bb^0_{(s)}$ mesons transforming into $B^0_{(s)}$ mesons. The condition to have \CP violation in the mixing is given by $|q/p| \neq 1$. However, to a very good approximation the SM predicts $|q/p| \simeq 1$, \emph{i.e.} \CP violation in the mixing is very small, as also confirmed by experimental determinations~\cite{HFAG,LHCb-PAPER-2013-033}. For neutral $B^0$ and $B^0_s$ mesons, the value of $|q/p|$ can be measured by means of the so-called semileptonic asymmetry
\begin{equation*}
A^{d(s)}_\mathrm{sl} \equiv \frac{\Gamma_{\Bb^0_{(s)}\rightarrow \ell^+X}(t) - \Gamma _{{B^0_{(s)}\rightarrow \ell^- X}}(t)}{\Gamma_{\Bb^0_{(s)}\rightarrow \ell^+X}(t) + \Gamma _{{B^0_{(s)}\rightarrow \ell^- X}}(t)} = \frac{1-|q/p|_{d(s)}^4}{1+|q/p|_{d(s)}^4},
\end{equation*}
which actually turns out to be not dependent on time.

Finally, \CP violation may arise in the interference between decay and mixing processes. Assuming \CPT invariance, the \CP asymmetry as a function of decay time for a neutral $B^0$ or $B^0_s$ meson decaying to a self-conjugate final state $f$, is given by
\begin{equation*}
A(t)\equiv\frac{\Gamma_{{\Bb}^0_{(s)} \to f}(t)-\Gamma_{B^0_{(s)} \to f}(t)}{\Gamma_{{\Bb}^0_{(s)} \to f}(t)+\Gamma_{B^0_{(s)} \to f}(t)}=\frac{-C_f \cos\left(\Delta m_{d(s)} t\right) + S_f \sin\left(\Delta m_{d(s)}t\right)}{\cosh\left(\frac{\Delta\Gamma_{d(s)}}{2} t\right) + A^{\Delta\Gamma}_f \sinh\left(\frac{\Delta\Gamma_{d(s)}}{2} t\right)}.
\end{equation*}
The quantities $C_f$, $S_f$ and $A^{\Delta\Gamma}_f$ are
\begin{equation*}
\begin{split}
C_{f} \equiv \frac{1-|\lambda_f|^2}{1+|\lambda_f|^2},\,\,\,\,\,\,\,\,\,\,\,S_{f} \equiv  \frac{2 \Im \lambda_f}{1+|\lambda_f|^2}\,\,\,\,\,\mathrm{and}\,\,\,\,\,\,A^{\Delta\Gamma}_f \equiv  - \frac{2 \Re \lambda_f}{1+|\lambda_f|^2},
\end{split}\label{eq:adirmix}
\end{equation*}
where $\lambda_f$ is given by
\begin{equation*}
\lambda_f \equiv \frac{q}{p}\frac{\bar{A}_f}{A_f}.
\end{equation*}
The parameter $\lambda_f$ is thus related to $B^0_{(s)}$--$\Bb^0_{(s)}$ mixing (via $q/p$) and to the decay amplitudes of the $B^0_{(s)} \to f$ decay ($A_f$) and of the $\Bb^0_{(s)} \to f$ decay ($\bar{A}_f$). With negligible $C\!P$ violation in the mixing ($|q/p|=1$), the terms $C_{f}$ and $S_{f}$ parametrise \CP violation in the decay and in the interference between mixing and decay, respectively.
The following relation between $C_{f}$, $S_{f}$ and $A^{\Delta\Gamma}_f$ holds
\begin{equation*}
\left( C_{f} \right)^2 + \left( S_{f} \right)^2 + \left( A^{\Delta\Gamma}_f \right)^2 = 1.
\end{equation*}
Notably, one has direct \CP violation ($\bar{A}_f \neq A_f$) when $C_{f} \neq 0$. But even in the case of suppressed \CP violation in the decay, it is still possible to observe \CP violation if a relative phase between $q/p$ and $\bar{A}_f / A_f $ exists. In such a case, one has $S_{f} = \Im (q/p \times \bar{A}_f / A_f )$. For example, for the $B^0 \to J/\psi K^0_S$ decay, to an approximation that is valid in the SM at the percent level or below, one has $S_{B^0 \to J/\psi K^0_S} = \sin(\phi_d)$ and $C_{B^0 \to J/\psi K^0_S} = 0$, with $\phi_d=2\beta$. Similarly, for the $B^0_s \to J/\psi \phi$ decay, \CP violation in the interference between mixing and decay gives access to $\phi_s=-2\beta_s$.

\subsection{Experimental determination of the unitarity triangle}

The experimental determination of the UT is here presented in brief. Many excellent reviews are available in the literature~\cite{HFAG,Ciuchini:2011ca,Agashe:2014kda,Bediaga:2012py,Fleischer:2002ys}, where detailed discussions on the various topics can be found. Several pieces of information must be combined by means of sophisticated fits in order to determine the apex of the UT with the highest possible precision. Relevant inputs to the fits are
\begin{itemize}
  \item $|\varepsilon_K|$: This parameter is determined by measuring indirect \CP violation in the neutral kaon mixing, using $K \to \pi \pi$, $K \to \pi l \nu$, and $K^0_L \to \pi^+\pi^- e^+ e^-$ decays, and provides a very important constraint on the position of the UT apex.
  \item $\left|V_{ub}\right|/\left|V_{cb}\right|$: The measurements of branching fractions of semileptonic decays governed by $b\rightarrow u l\bar{\nu}$ and $b\rightarrow c l\bar{\nu}$ transitions give information about the magnitudes of $V_{ub}$ and $V_{cb}$, respectively. The ratio between these two quantities constrains the side of the UT between the $\gamma$ and $\alpha$ angles. 
  \item $\Delta m_{d}$: This parameter represents the frequency of $B^{0}-\bar{B}^{0}$ mixing. It is proportional to the magnitude of $V_{td}$ and thus constrains the side of the UT between the $\beta$ and $\alpha$ angles. 
  \item $\Delta m_{s}/\Delta m_{d}$: $\Delta m_{s}$ is the analogue of $\Delta m_{d}$ in the case of $B_{s}^{0}-\bar{B}_{s}^{0}$ mixing and its value is proportional to the magnitude $V_{ts}$. However, in order to reduce theoretical uncertainties on hadronic parameters determined using Lattice QCD calculations, the use of the ratio $\Delta m_{s}/\Delta m_{d}$ is more effective. This also provides a constraint on the side of the UT between the $\beta$ and $\alpha$ angles.
  \item $\sin(2\beta)$: This quantity is mainly determined from time-dependent \CP-violation measurements of $B^{0}\rightarrow J/\psi K^{0}_S$ decays, and provides a powerful constraint on the angle $\beta$ of the UT.
  \item $\alpha$: This UT angle is determined from the measurements of \CP-violating asymmetries and branching fractions in $B\rightarrow\pi\pi$, $B\rightarrow\rho\rho$ and $B\rightarrow\rho\pi$ decays.
  \item $\gamma$: The determination of this angle is performed by measuring time-integrated \CP-violating asymmetries and branching fractions of $B\rightarrow D^{(*)}K^{(*)}$ decays, and time-dependent \CP violation in $B^0_s \to D_s^\pm K^\mp$ decays. 
\end{itemize}

\begin{figure}[tb]
  \begin{center}
    \includegraphics[width=0.48\textwidth]{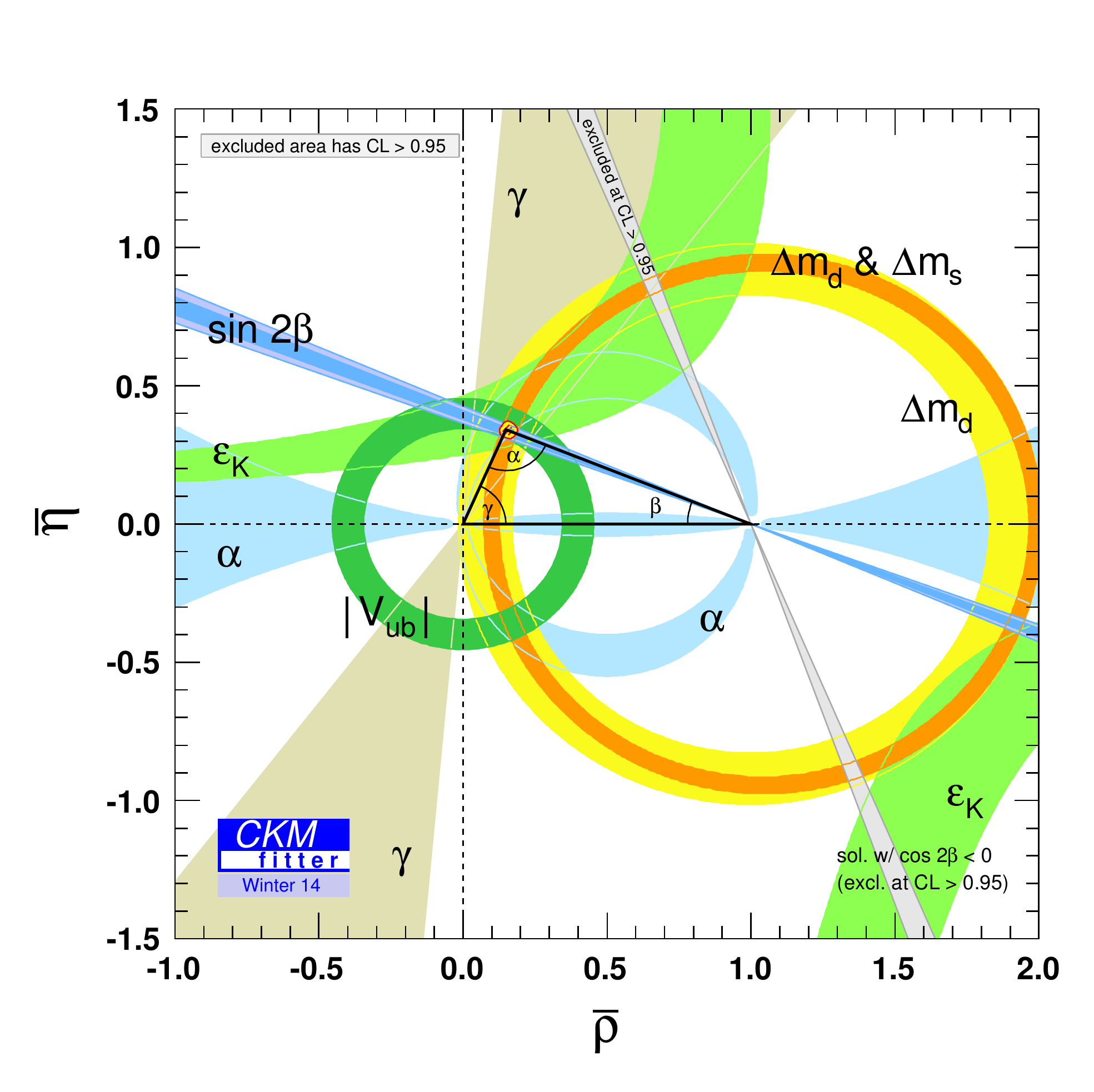}
    \includegraphics[width=0.48\textwidth]{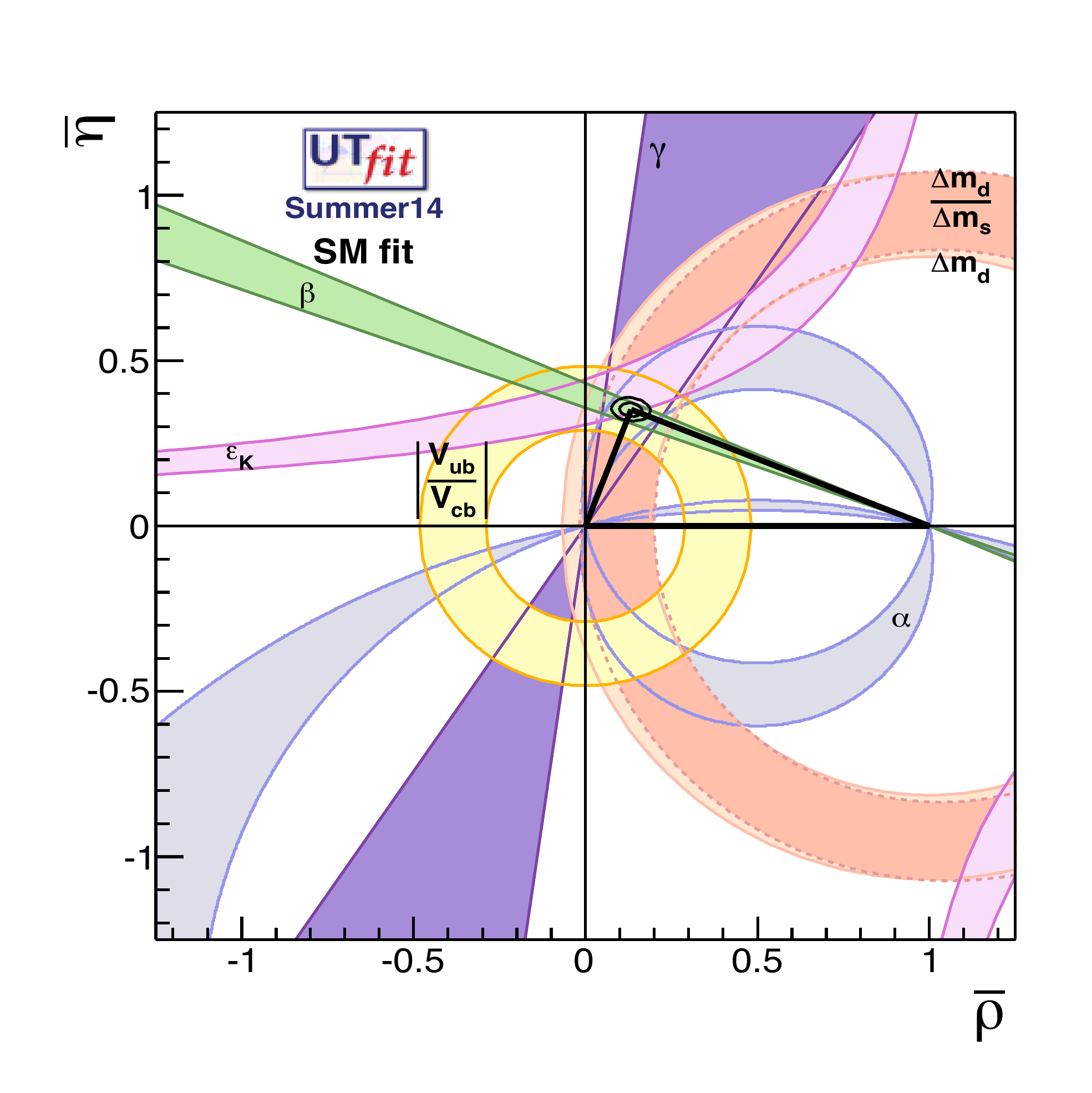}
  \end{center}
  \caption{Results of global CKM fits performed by the (left) CKMfitter~\cite{Charles:2015gya} and (right) UTfit~\cite{Bona:2006ah} groups. The 95\% probability regions corresponding to each of the experimental measurements are indicated by filled areas of different colours. The various areas intersect in the position of the UT apex.}\label{fig:utFit}
\end{figure}

\begin{table}[tb]
\caption{Results of global CKM fits by the CKMfitter~\cite{Charles:2015gya} and UTfit~\cite{Bona:2006ah} groups.}
\begin{center}
\begin{tabular}{@{}lcccc@{}} 
Group        & $A$ & $\lambda$ & $\bar{\rho}$ & $\bar{\eta}$ \\
\hline
CKMfitter \bigstrut  & $0.810^{\:+\: 0.018}_{\:-\: 0.024}$ & $0.22548^{\:+\: 0.00068}_{\:-\: 0.00034}$ & $0.1453^{\:+\: 0.0133}_{\:-\: 0.0073}$ & $0.343^{\:+\: 0.011}_{\:-\: 0.012}$ \\
UTfit          & $0.821 \pm 0.012$ & $0.22534 \pm 0.00065$ & $0.132 \pm 0.023$ & $0.352 \pm 0.014$ \\ 
\end{tabular}
\end{center}
\label{tab:utFitParameter}
\end{table}

World averages of the various experimental measurements are kept up to date by the Heavy Flavour Averaging Group~\cite{HFAG}. Each of these measurements yields a constraint on the position of the UT apex, \emph{i.e.} on the values of the $\bar{\rho}$ and $\bar{\eta}$ parameters. Two independent groups, namely CKMfitter~\cite{Charles:2015gya} and UTfit~\cite{Bona:2006ah}, regularly perform global CKM fits starting from the same set of experimental and theoretical inputs, but using different statistical approaches. In particular, CKMfitter performs a frequentist analysis, whereas UTfit follows a Bayesian method. Their latest results are displayed in Fig.~\ref{fig:utFit}. Each of the experimental constraints is represented as a 95\% probability region by a filled area of different colour. The intersection of all regions identifies the position of the UT apex. The level of agreement amongst the various regions in pinpointing the UT apex is also a measure of the level of consistency of the KM mechanism with data. If (at least) one of the areas were not in agreement with the others, that would be an indication of the existence of BSM physics. At present, no striking evidence of any disagreement is found. The latest values of $A$, $\lambda$, $\bar{\rho}$ and $\bar{\eta}$ obtained by the CKMfitter and UTfit groups are reported in Table~\ref{tab:utFitParameter}. Besides small fluctuations, also stemming from slightly different theoretical inputs and statistical procedures, the two sets of results are in good agreement.

\section{Overview of beauty physics at the LHC}\label{Sec:BPhysLHC}
\subsection{\CP violation}\label{Sec:CPVBeauty}
Owing to the legacy of the \B factories~\cite{Bevan:2014iga}, we have entered the era of precision tests of CKM physics, where ultimate sensitivity is needed to search for new sources of \CP violation beyond the single phase of the CKM matrix.

\begin{figure}[tb]
  \begin{center}
\includegraphics[width=0.8\textwidth]{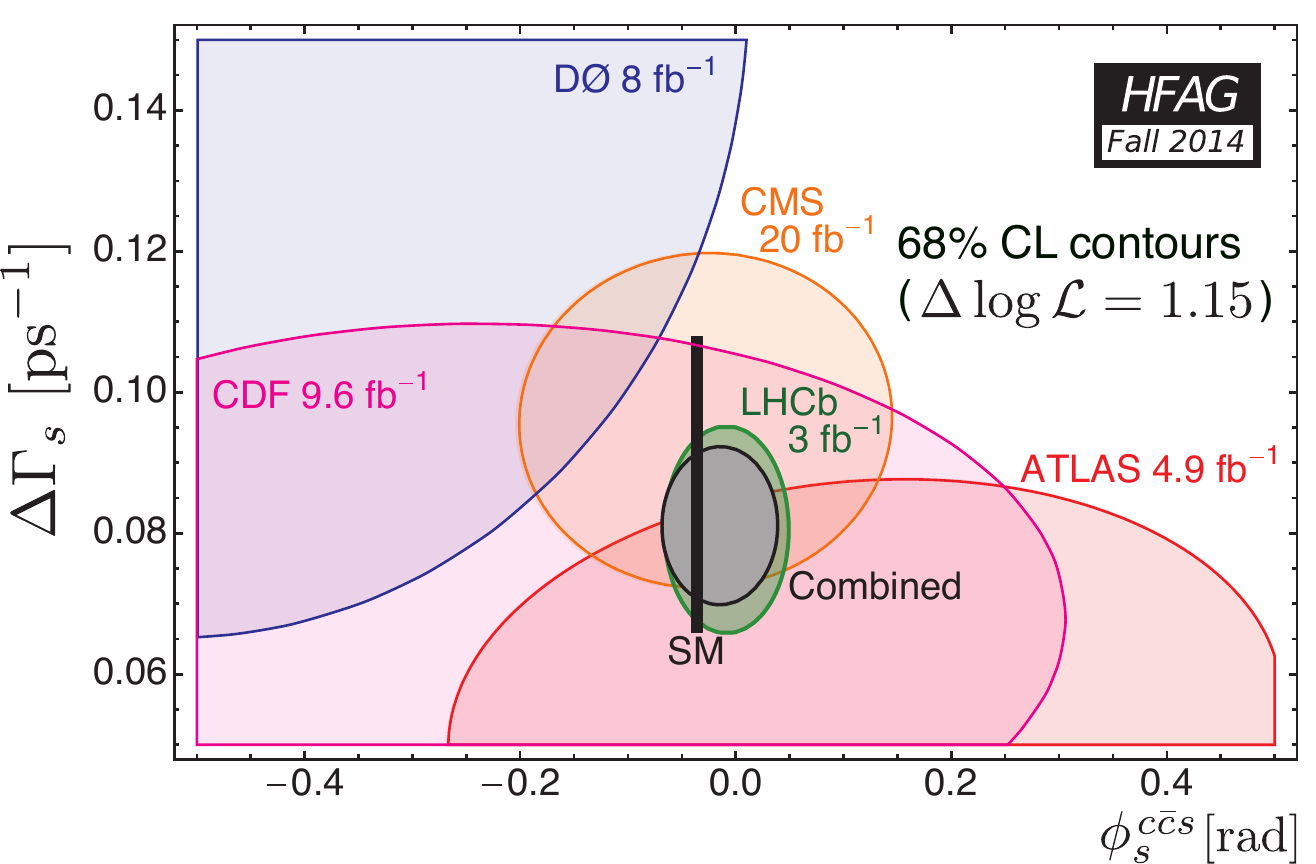}\hskip 0.02\textwidth
  \end{center}
\caption{Constraints on $\Delta\Gamma_s$ and $\varphi_s^{\cquark\cquarkbar\squark}$ from various experiments.}
\label{Fig:2014-059}
\end{figure}

The LHC is often considered as a \Bs-meson factory, due to the large production cross-section and to the excellent capabilities of the LHC experiments to precisely resolve \Bs oscillations. This has opened the door to precision measurements of the \CP-violating phase $\varphi_s^{\cquark\cquarkbar\squark}$, which is equal to $-2\beta_s$ in the SM, neglecting sub-leading penguin contributions.
It has been measured at the LHC by ATLAS, CMS and LHCb using the flavour eigenstate decays \decay{\Bs}{\jpsi\Kp\Km}~\cite{Aad:2014cqa,CMS-PAS-BPH-13-012,LHCb-PAPER-2013-002} and 
\decay{\Bs}{\jpsi\pip\pim}~\cite{LHCb-PAPER-2014-019}. Recently LHCb has used the decay 
\decay{\Bs}{\jpsi\Kp\Km} for the first time in a polarisation-dependent way~\cite{LHCB-PAPER-2014-059}. 
The quantity $\varphi_s^{\cquark\cquarkbar\squark}$ has also been measured with a fully hadronic final state using the decay
\decay{\Bs}{\Dsp\Dsm} with \decay{\Dspm}{\Kp\Km\pipm}, yielding $0.02\pm0.17\pm0.02$~rad~\cite{LHCb-PAPER-2014-051}. 
Combining all determinations, LHCb obtains $\varphi_s^{\cquark\cquarkbar\squark}=-0.010\pm 0.039$~rad. Including also other results, and in particular those from ATLAS and CMS, the uncertainty is further reduced, obtaining $\varphi_s^{\cquark\cquarkbar\squark}=-0.015\pm 0.035$~rad.
The  constraints on $\varphi_s^{\cquark\cquarkbar\squark}$ and on the decay width difference 
$\Delta\Gamma_s$  are shown in Fig.~\ref{Fig:2014-059}, together with the corresponding SM expectations.

With the precision reaching the degree level, the effects of suppressed penguin topologies cannot be neglected anymore~\cite{Fleischer:1999nz,Fleischer:1999zi,Fleischer:1999sj,Faller:2008zc,DeBruyn:2010hh,Ciuchini:2005mg}.
Such effects may lead to a shift $\delta \varphi_s$ in the measured value of $\varphi_s^{\cquark\cquarkbar\squark}$, which can be constrained using Cabibbo-suppressed decay modes, where penguin topologies are relatively more prominent. 
This programme has started with studies of the decays 
\decay{\Bs}{\jpsi\KS}~\cite{LHCb-PAPER-2013-015,LHCb-PAPER-2015-005},
\decay{\Bs}{\jpsi\Kstarzb}~\cite{LHCb-PAPER-2012-014} and more recently 
\decay{\Bz}{\jpsi\pip\pim}~\cite{LHCb-PAPER-2014-058}. These studies enable a constraint on $\delta \varphi_s$ to be placed in the range $[-0.018,0.021]$~rad at 68\% CL. Considering the present 
uncertainty of $0.039$~rad, such a shift needs to be to constrained further.

Another interesting test of the SM is provided by the measurement of the mixing phase
$\varphi_s^{\squark\squarkbar\squark}$
with a penguin-dominated mode as \decay{\Bs}{\phi\phi}.
In this case the measured value is $-0.17\pm 0.15\pm 0.03$~rad~\cite{LHCb-PAPER-2014-026},
which is compatible with the SM expectation.

Similarly, the decays \decay{\B}{hh} with $h=\pion,\kaon$ receive large contributions from 
penguin topologies, and are sensitive to $\gamma$ and $\beta_s$.
LHCb for the first time measured time-dependent \CP-violating observables in
\Bs decays using the decay 
\decay{\Bs}{\Kp\Km}~\cite{LHCb-PAPER-2013-040}. Using methods outlined in 
Refs.~\cite{Fleischer:1999pa,Fleischer:2007hj,Ciuchini:2005mg}, a combination of this
and other results from $B \to \pi\pi$ modes enables the determination
$-2\beta_s = -0.12 \aerr{0.14}{0.16}$\:\rad 
using as input the angle $\gamma$ from tree-level decays (see below), or
$\gamma= (63.5\aerr{7.2}{6.7})^\circ$ relating $-2\beta_s$ to the 
SM expectation~\cite{LHCb-PAPER-2013-045}.
These values are in principle sensitive to
the amount of U-spin (a subgroup of $SU(3)$ analogous to isospin, but involving $d$ and $s$ quarks instead of $d$ and $u$ quarks) 
breaking in the involved decay amplitudes and are given here 
for a maximum allowed breaking of 50\%. 
This value of $\gamma$ can be compared to that obtained from tree-dominated
\decay{\B}{\D\kaon} decays, where the \CP-violating phase appears in the interference of
the \decay{\bquark}{\cquark} and \decay{\bquark}{\uquark} topologies. The determination of $\gamma$
from tree decays is considered free from contributions beyond the SM
and unaffected by hadronic uncertainties. Yet, its precise measurement is important
to test the consistency of the KM mechanism, also allowing for comparisons with measurements
from modes dominated by penguin topologies.

The most precise determination of $\gamma$ from a single tree-level decay mode is achieved with the decay
\decay{\Bp}{D\Kp} followed by \decay{D}{\KS h^+h^-} with 
$h=\pion,\kaon$~\cite{LHCb-PAPER-2014-041}, yielding $\gamma=(62\aerr{15}{14})\degrees$. 
Here the interference of the \Dz and \Dzb decays to $\KS h^+h^-$ 
is exploited to measure \CP asymmetries~\cite{Giri:2003ty}. The method needs external
input in the form of a measurement of the strong phase over the Dalitz plane
of the \D decay, coming from CLEO-c data~\cite{Libby:2010nu}. The same decay mode
is also used in a model-dependent measurement~\cite{LHCb-PAPER-2014-017}.

A different way for determining $\gamma$ is provided by the decay 
\decay{\Bs}{D_s^\pm\kaon^\mp}~\cite{Dunietz:1987bv,Aleksan:1991nh,Fleischer:2003yb,LHCb-PAPER-2014-038}. In this case the phase 
is measured in a time-dependent tagged \CP-violation analysis. Using a dataset 
corresponding to 1\:\invfb, LHCb determines $\gamma=(115\aerr{28}{43})\degrees$,
which is not yet competitive with other methods but will provide important cross-checks
 with more data.

\begin{figure}[tb]
  \begin{center}
\includegraphics[width=0.8\textwidth]{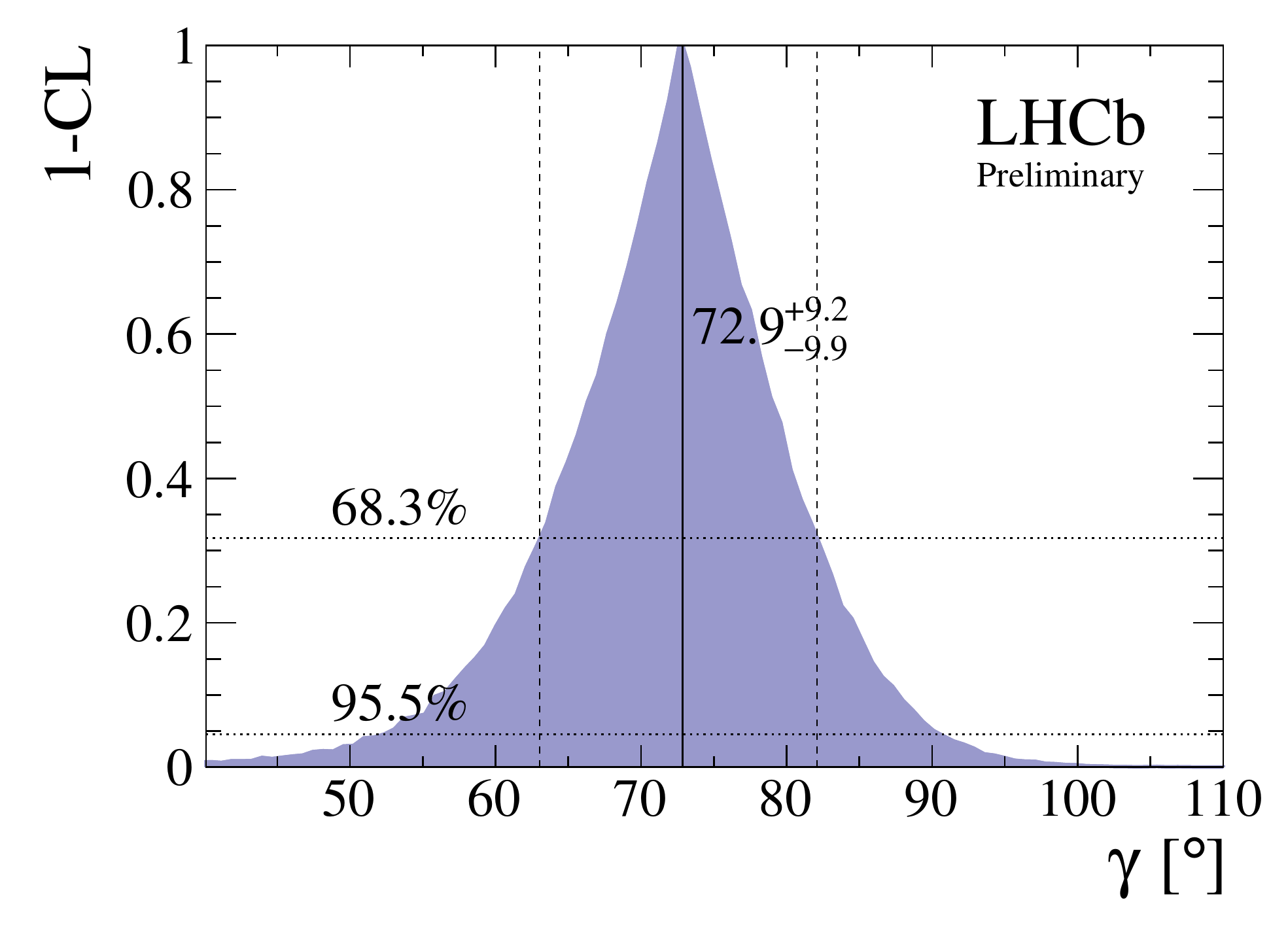}
  \end{center}
\caption{LHCb combination of \decay{\B}{D\kaon} decays measuring $\gamma$~\cite{LHCb-CONF-2014-004}.}\label{Fig:CONF-2014-004}
\end{figure}

The $\gamma$ measurements of 
Refs.~\cite{LHCb-PAPER-2014-041,LHCb-PAPER-2014-038,LHCb-PAPER-2012-001,LHCb-PAPER-2012-055,LHCb-PAPER-2013-068,LHCb-PAPER-2014-028} are 
then combined in an LHCb average~\cite{LHCb-CONF-2014-004}.
Using all \decay{\B}{D\kaon} decay modes one finds $\gamma=(73\aerr{9}{10})\degrees$,
which is more precise than the corresponding combination of measurements from the \B factories~\cite{Bevan:2014iga}. The LHCb likelihood profile is shown in Fig.~\ref{Fig:CONF-2014-004}.

\begin{figure}[tb]
  \begin{center}
\includegraphics[width=0.7\textwidth]{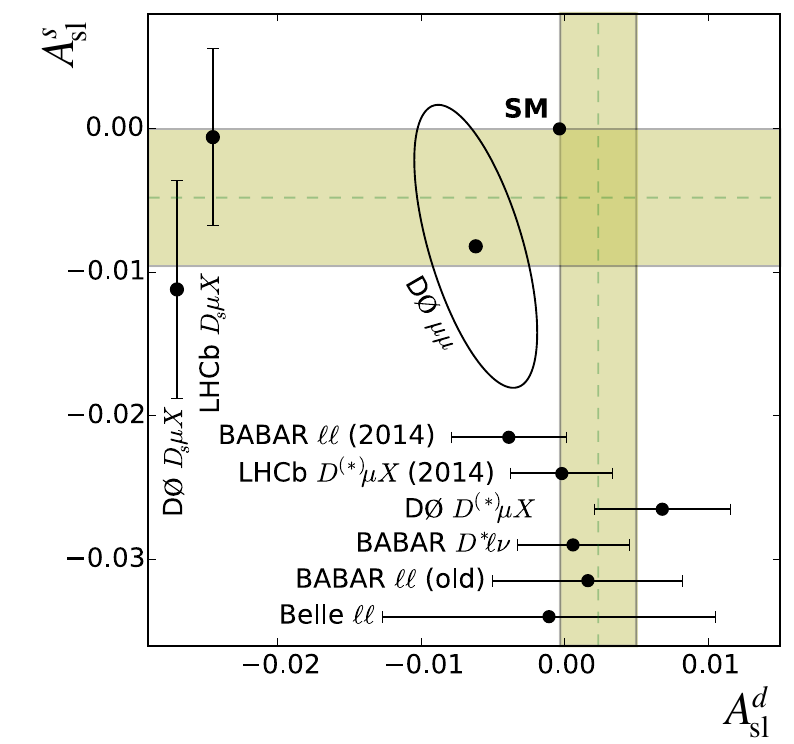}
  \end{center}
\caption{Experimental constraints on $A_\text{sl}^d$ and $A_\text{sl}^s$ from various experiments. The symbol $\ell$ stands for electron or muon. The horizontal and vertical bands represent the uncertainties on the averages of the experimental measurements. The elliptic contour represents the measurement of the same-sign dimuon asymmetry by the D0 experiment.}\label{Fig:2014-053}
\end{figure}

The same-sign dimuon asymmetry measured by the \dzero collaboration~\cite{Abazov:2013uma} 
and interpreted as a combination of the semileptonic asymmetries $A_\text{sl}^d$ and $A_\text{sl}^s$ in \Bd and \Bs decays, respectively, differs from the SM expectation by $3\sigma$. So far LHCb has not been able to confirm
or disprove this result. The measurement from LHCb looks at the \CP asymmetry
between partially reconstructed \decay{\B}{\D\mu\nu} decays, where the flavour of the 
\D meson identifies that of the \B. The measured value of 
$A_\text{sl}^s$~\cite{LHCb-PAPER-2013-033} and the
newly reported $A_\text{sl}^d$~\cite{LHCb-PAPER-2014-053} are both consistent 
with the SM and the \dzero values. The world average including measurements from the
\B factories and \dzero is not more conclusive, as shown in Fig.~\ref{Fig:2014-053}.

Large \CP violation has been found in charmless \bquark-hadron decays 
like \decay{\Bp}{h^+h^-h^+}~\cite{LHCb-PAPER-2014-044} ($h=\pion,\kaon$) and 
\decay{\Bp}{\proton\antiproton h^+}~\cite{LHCb-PAPER-2014-034}. Particularly striking 
features of these decays are the very large asymmetries observed in small
regions of the phase-space not related to any resonance, 
which are opposite in sign for \decay{\Bpm}{h^\pm\Kp\Km}
and \decay{\Bpm}{h^\pm\pip\pim} decays. This observation could be a sign of long-distance
$\pip\pim\leftrightarrow\Kp\Km$ rescattering.

Another important field is the study of \CP violation in beauty baryons.
The probability that a $b$ quark hadronises to a \Lb baryon is measured to be surprisingly large at the 
LHC in the forward region~\cite{LHCb-PAPER-2014-004}, almost half of that
to a \Bd meson. These baryons can be used for measurements of \CP violation
with better precision than \Bs mesons. Searches have been performed by LHCb
with the decays \decay{\Lb}{\jpsi\proton\pim}~\cite{LHCb-PAPER-2014-020},
\decay{\Lb}{\Kz\proton\pim}~\cite{LHCb-PAPER-2013-061},
and by CDF with \decay{\Lb}{\proton \pi^-} and \decay{\Lb}{\proton K^-}~\cite{Aaltonen:2011qt}.
It is worth noting that no evidence for \CP violation in any
decay of a baryon has ever been reported to date.

\subsection{Rare electroweak decays}\label{Sec:blls}
The family of decays \decay{\bquark}{\squark\ellp\ellm} is a laboratory for BSM-physics searches on its own. In particular the exclusive decay \decay{\Bz}{\Kstarz\ellp\ellm} ($\ell=e,\mu$) provides a very rich set of observables with different sensitivities to BSM physics and for which the available SM predictions are affected by varying levels of hadronic uncertainties. For some ratios of such observables, most of the theoretical uncertainties cancel out, thus providing a clean test of the SM~\cite{Ali:1999mm,Kruger:2005ep,Altmannshofer:2008dz,Egede:2008uy,Bobeth:2008ij,Descotes-Genon:2013vna}. 

The differential decay width with respect to the dilepton mass squared $q^2$, 
the forward-backward asymmetry $A_\text{FB}$, and the longitudinal polarisation fraction $F_\text{L}$ of the \Kstar resonance have been measured by many experiments~\cite{Aubert:2006vb,Wei:2009zv,Aaltonen:2011ja,Chatrchyan:2013cda,ATLAS-CONF-2013-038,LHCb-PAPER-2013-019} with no significant sign of deviations
from the SM expectation.

In a second analysis of the already published 2011 data~\cite{LHCb-PAPER-2013-019}, 
LHCb has published another set of angular observables~\cite{LHCb-PAPER-2013-037}
suggested by Ref.~\cite{Descotes-Genon:2013vna}. In particular a $3.7\sigma$
local deviation from the SM expectation of one of these observables has been found in one bin of $q^2$. 
This measurement has triggered a lot of interest in the theoretical community, with interpretation articles being submitted very quickly to journals. See Refs.~\cite{Gauld:2013qja,Descotes-Genon:2013wba,Altmannshofer:2013foa,Datta:2013kja,Mahmoudi:2014mja} for a small subset. It is not clear whether this discrepancy is an experimental fluctuation, is due to under-estimated form factor uncertainties (see Ref.~\cite{Beaujean:2013soa}), or is the manifestation of a heavy $Z'$ boson, among many other suggested explanations. The contribution
of \cquark{}\cquarkbar resonances is also being questioned~\cite{Lyon:2014hpa}
after the LHCb observation of \decay{\Bp}{\psi(4160)\Kp} with 
\decay{\psi(4160)}{\mumu}~\cite{LHCb-PAPER-2013-039}, 
where the $\psi(4160)$ and its interference with the non-resonant component 
accounts for 20\% of the rate for dimuon masses 
above 3770\:\mevcc. Such a large contribution was not expected.

Given a hint of abnormal angular distributions, LHCb tried to look for 
other deviations in several asymmetry measurements. The \CP asymmetry in 
\decay{\Bd}{K^{(*)0}\mumu} and \decay{\Bpm}{\Kpm\mumu} decays turns out to be compatible with zero as expected~\cite{LHCb-PAPER-2014-032}, as does the isospin asymmetry between \decay{\Bd}{K^{(*)0}\mumu}
and \decay{\Bu}{K^{(*)+}\mumu} decays~\cite{LHCb-PAPER-2014-006}. The lepton universality factor
$R_K=\frac{{\cal B}(\decay{\Bp}{\Kp\mumu})}{{\cal B}(\decay{\Bp}{\Kp\epem})}$
is measured to be $0.745\aerr{0.090}{0.074}\pm0.036$~\cite{LHCb-PAPER-2014-024} in the $1<q^2<6\:\gevgevcccc$ range, which indicates a $2.6\sigma$ deviation from unity. 
This result can be interpreted as a possible indication of a new vector particle that would couple
more strongly to muons and interfere destructively with the SM vector 
current~\cite{Hiller:2014yaa,Ghosh:2014awa,Altmannshofer:2014rta,Crivellin:2015mga,Glashow:2014iga}.

\subsection{Observation of the $B^0_s \to \mu^+\mu^-$ decay}
The measurement of the branching fractions of the rare $B^0 \to  \mu^+\mu^-$ and $B^0_s \to  \mu^+\mu^-$ decays is considered amongst the most promising avenues to search for BSM effects at the LHC. These decays proceed via FCNC processes and are highly suppressed in the SM. Moreover, the helicity suppression of axial vector terms makes them sensitive to scalar and pseudoscalar BSM contributions that can alter their branching fractions with respect to SM expectations. The untagged time-integrated SM predictions for the branching fractions of these decays are~\cite{Bobeth:2013uxa}
\begin{eqnarray*}
\mathcal{B}(\bsmumu)_{SM} = (3.66\pm0.23)\times 10^{-9} \nonumber \,,\\
\mathcal{B}(\bdmumu)_{SM} = (1.06\pm0.09)\times 10^{-10} \nonumber \,,
\end{eqnarray*}
which are obtained using the latest combination of values for the top-quark mass from LHC and Tevatron experiments~\cite{ATLAS:2014wva}.
The ratio \RB between these two branching fractions is also a powerful tool to discriminate amongst BSM models. In the SM it is predicted to be
\begin{eqnarray*}
\RB = \frac{\BRof{\bdmumu}}{\BRof{\bsmumu}}   
= \frac{\tau_{\bd}}{1/\Gamma_H^{s}} \left(\frac{f_{\bd}}{f_{\bs}}\right)^2 
 \left|\frac{V_{td}}{V_{ts}}\right|^2 \frac{M_{\bd} \sqrt{1 - \tfrac{4 
 m_{\mu}^2}{M_{\bd}^2}}}{M_{\bs} \sqrt{1 - \tfrac{4 m_{\mu}^2}{M_{\bs}^2}}}
= \RBSM\,, 
\end{eqnarray*}
where $\tau_{\bd}$ and $1/\Gamma_H^{s}$ are the lifetimes of the \bd and of the heavy mass eigenstate 
of the $B^0_{s}$--$\Bb^0_{s}$  system, $M_{\bds}$ is the mass and $f_{\bds}$ is the decay constant of the \bds meson, $V_{td}$ and $V_{ts}$ are the elements of the CKM matrix and $m_{\mu}$ is the mass of the muon. In minimal flavour-violating BSM scenarios, the branching fractions of both decays can change, but their ratio is predicted to be equal to that of the SM.

The LHCb collaboration reported the first evidence of the \bsmumu decay with a $3.5\,\sigma$ 
significance in 2012 using 2\fb\ of data~\cite{LHCb-PAPER-2012-043}. One year later, CMS and LHCb presented their updated results based on 25\fb and 3\fb, respectively~\cite{LHCb-PAPER-2013-046,Chatrchyan:2013bka}. 
The two measurements resulted in good agreement with comparable precisions. However, none of them was precise enough to claim the first observation of the \bsmumu decay. A na\"ive combination of \cms and \lhcb results was presented in 2013~\cite{LHCb-CONF-2013-012}, but no attempt was made to take into account all correlations stemming from common physical quantities, and no statistical significance for the existence of the signals was provided.

More recently, a combination of the CMS and LHCb results based on a simultaneous fit to the two datasets has been performed. This fit correctly takes into account correlations between the input parameters. The \cms and \lhcb experiments have very similar analysis strategies. \bdsmumu candidates are selected as two oppositely charged tracks. A soft first selection is applied   
in order to reduce the background while keeping high the efficiency on the signal.
After this selection, the remaining backgrounds are mainly due to random combinations of muons from semileptonic $B$ decays (combinatorial background), semileptonic decays, such as \Bhmunu, \Bhmumu and \Lbpmunu, and \Bhhprime decays (peaking background) where hadrons are misidentified as muons. 
Further separation between signal and background is achieved exploiting the power of a multivariate classifier. The classification of the events is done using the dimuon invariant mass \mmumu and the multivariate classifier output. 
The multivariate classifier is trained using kinematic and geometrical variables. 
The calibration of the dimuon mass \mmumu is performed using the dimuon resonances and, for \lhcb, also by using \Bhhprime decays.
For both analyses the \bdsmumu yield is normalised with respect to the \bujpsik yield, taking into account the hadronisation fractions of a $b$ quark to \bs and \bd mesons measured by the \lhcb experiment~\cite{LHCb-PAPER-2011-006,LHCb-PAPER-2011-018,LHCb-CONF-2013-011}. 
\lhcb also uses the \bdkpi decay as a normalisation channel.

\begin{figure}[tb]
\centering
\includegraphics[width=0.8\textwidth]{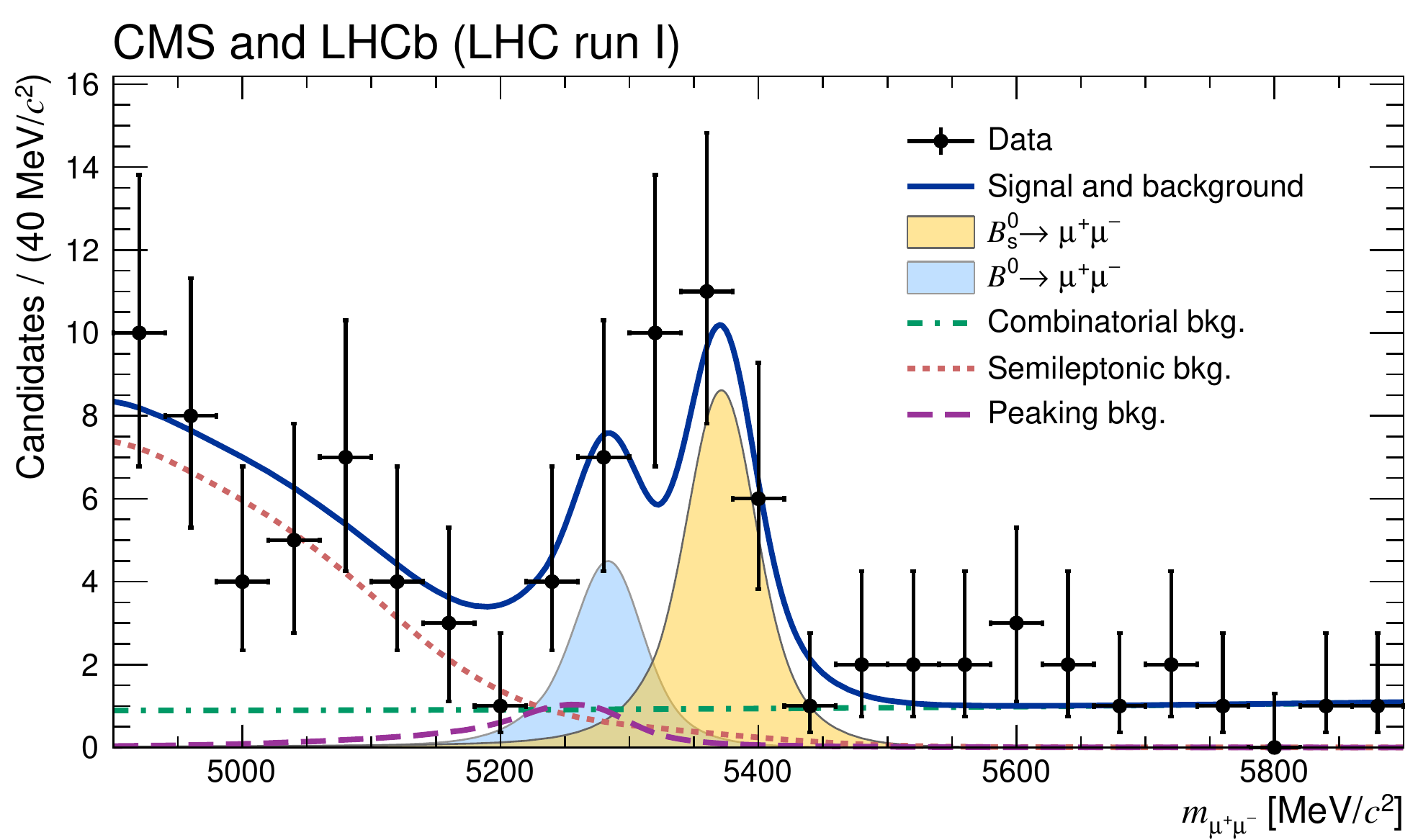}
\caption{Dimuon mass distribution for the six multivariate-classifier categories with highest \bs signal purity. The result of the simultaneous fit is overlaid.}
\label{fig1}
\end{figure}

\begin{figure}[tb]
\centering
\includegraphics[width=1\textwidth]{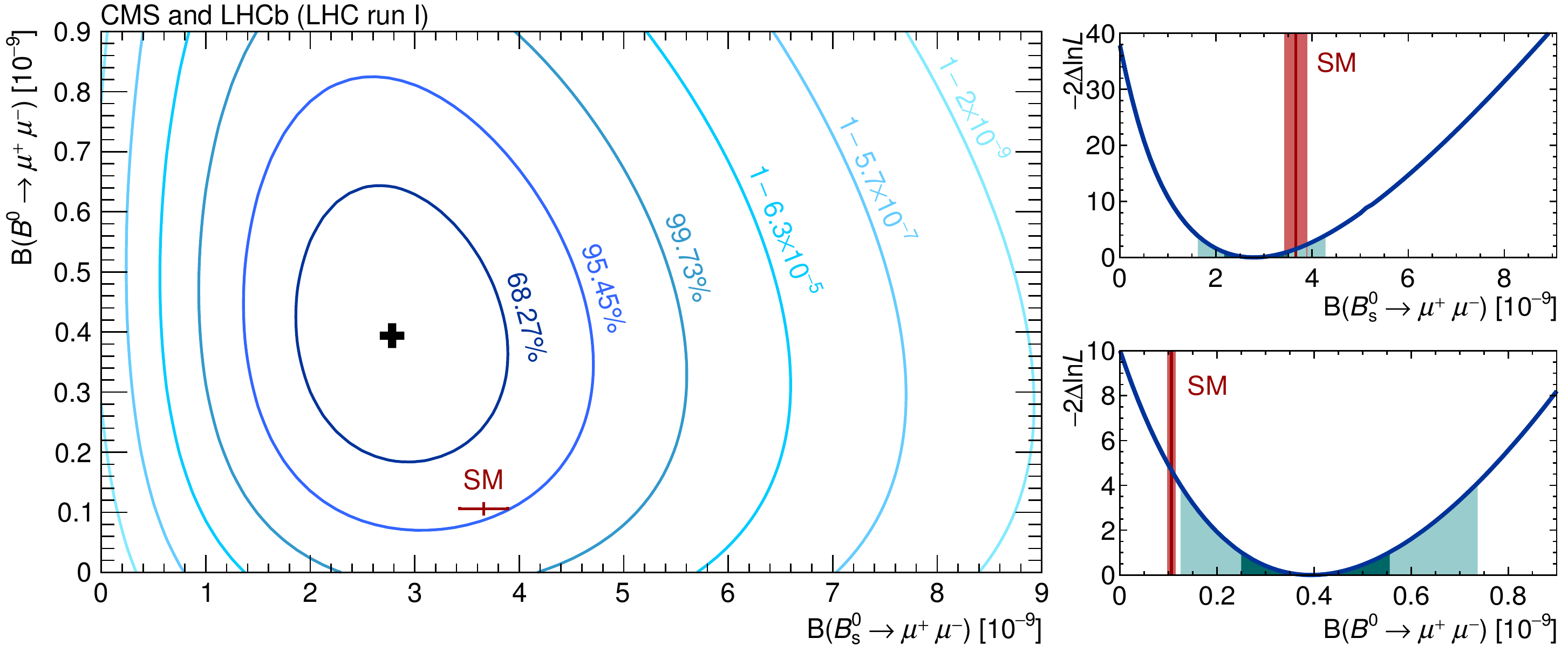}
\caption{(left) Likelihood contours in the \BRof\bsmumu--\BRof\bdmumu plane. Likelihood profile for (top-right) \BRof\bsmumu and (bottom-right) \BRof\bdmumu. The dark and light areas define the $\pm 1\,\sigma$ and $\pm 2\,\sigma$ confidence intervals, respectively. The SM expectations are indicated with vertical bands.}
\label{fig2}
\end{figure}

\begin{figure}[tb]
\centering
\includegraphics[width=0.8\textwidth]{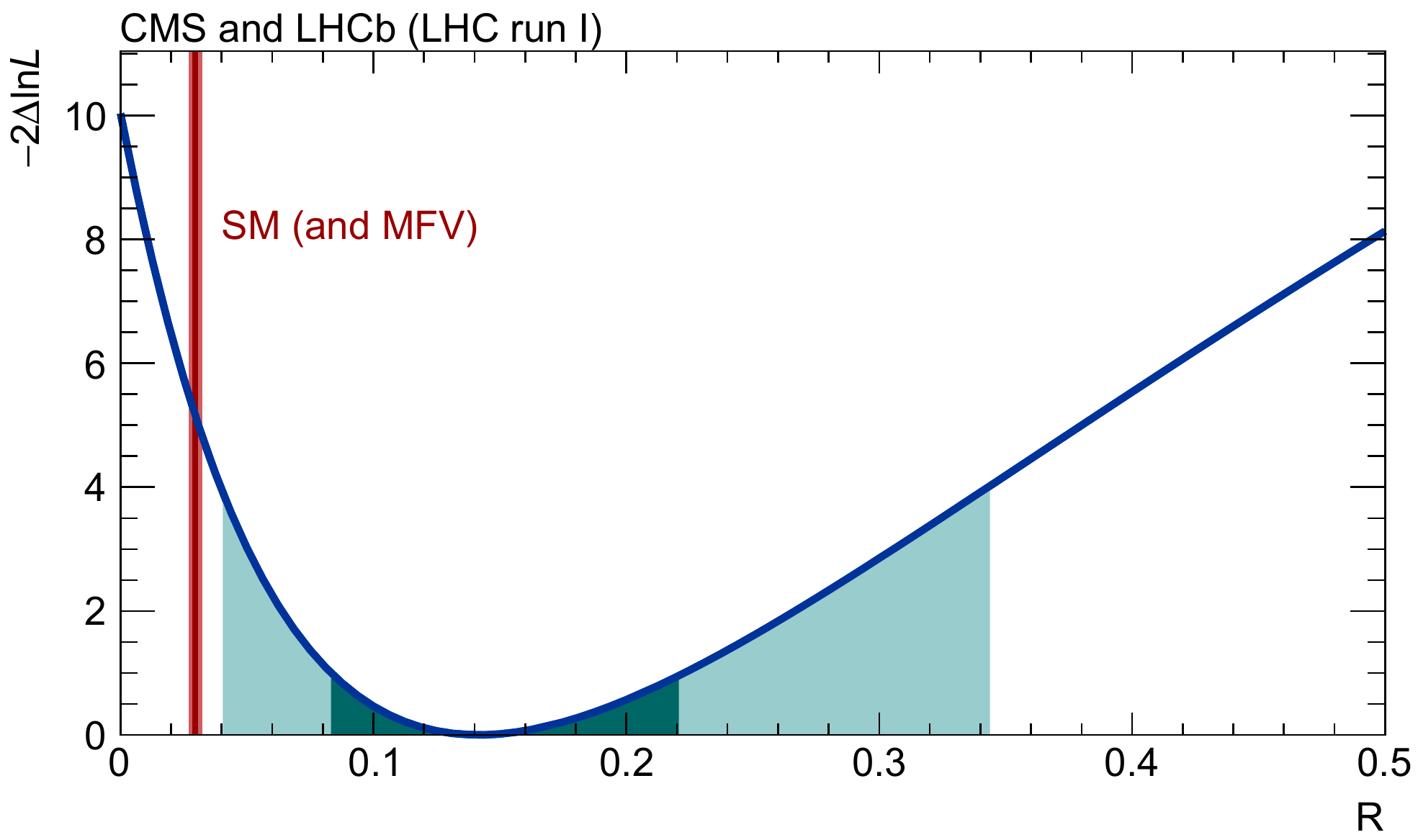}
\caption{Likelihood profile for \RB. The dark and light areas define the $\pm 1\,\sigma$ and $\pm 2\,\sigma$ confidence intervals, respectively. The SM expectation is indicated with a vertical band.}
\label{fig3}
\end{figure}

A simultaneous fit is performed to evaluate the branching fractions of the \bsmumu and \bdmumu decays. The CMS and LHCb datasets are used together as in a single combined experiment. 
A simultaneous unbinned extended maximum likelihood fit is performed to the invariant mass spectrum in 20 categories of multivariate
classifier output for the two experiments: 8 categories for \lhcb and 12 categories for \cms. The various categories are characterised by construction by different values of signal purity. 
In each category the mass spectrum is described as the sum of each background source and the two signals. The parameters shared between the two experiments are the branching fractions of the two signal decays being looked for, \BRof{\bsmumu} and \BRof{\bdmumu}, the already measured branching fraction of the common normalisation channel \BRof{\bujpsik}, and the ratio of the hadronisation fractions \fsfd.
Assuming the SM, $94\pm7$ \bsmumu events and $10.5\pm0.6$ \bdmumu events are expected in the full dataset.
For illustrative purposes, Fig.~\ref{fig1} shows the dimuon mass distribution for the events corresponding to the six multivariate categories with highest $B^0_s$ signal purity. The results of the simultaneous fit are~\cite{LHCb-PAPER-2014-049}
\begin{eqnarray*}
\br[\bsmumu]{\bsmumumeas} \,, \nonumber \\
\br[\bdmumu]{\bdmumumeas} \,. \nonumber
\end{eqnarray*}
The statistical significances, evaluated using the Wilks' theorem~\cite{Wilks}, are $6.2\,\sigma$ and $3.2\,\sigma$ for \bsmumu and \bdmumu, respectively. The expected significances assuming the SM branching fractions are $7.4\,\sigma$ and $0.8\,\sigma$ for \bs and \bd channels, respectively.
Since the Wilks' theorem shows a \bdmumu signal significance slightly above the $3\,\sigma$ level, a more refined method based on the Feldman-Cousins construction~\cite{prd573873} is also used for the \bdmumu mode. A statistical significance of $3.0\,\sigma$ is obtained in this case. The Feldman-Cousins confidence intervals at $\pm 1\,\sigma$ and $\pm 2\,\sigma$ are $[2.5,5.6]\times10^{-10}$ and $[1.4,7.4]\times10^{-10}$, respectively.
In Fig.~\ref{fig2} the likelihood contours in the \BRof\bsmumu--\BRof\bdmumu plane are shown. In the same figure, the likelihood profile for each signal mode is displayed.
The compatibility of \BRof{\bsmumu} and \BRof{\bdmumu} with the SM is evaluated at the $1.2\,\sigma$ and $2.2\,\sigma$ levels, respectively.
A separate fit to the ratio of \bd to \bs gives $\RB = 0.14^{\:+\:0.08}_{\:-\:0.06}$, which is compatible with the SM at the $2.3\,\sigma$ level. The likelihood profile for \RB is shown in Fig.~\ref{fig3}.

\section{Conclusions}

The LHC is the new \bquark-hadron factory. The ATLAS, CMS and LHCb experiments will be dominating heavy-flavour physics in the next decade, together with the forthcoming Belle II experiment, and even beyond with the high luminosity LHC phase. During Run I, ATLAS, CMS and LHCb have performed fundamental measurements in the field of \CP violation and rare decays of $B$ mesons. In this paper we have discussed some of these measurements, notably including that of the $B^0_s$-meson mixing phase from $b\to c\bar{c}s$ decays, $\varphi_s^{c\bar{c}s}$; of the UT angle $\gamma$ from tree-level decays; of $B^0$ and $B^0_s$ semileptonic asymmetries; of angular observables in $b \to s \ellp\ellm$ transitions; and of the $B^0_s \to \mu^+\mu^-$ branching fraction, with the first observation of this decay at more than five standard deviations. No striking evidences of deviations from SM expectations have emerged from any of these measurements so far. However, they have enabled strong constraints to be set on many BSM models. The upcoming Run II, with its higher centre-of-mass energy translating into a higher \bquark{}\bquarkbar cross-section, will witness substantial improvements in the study of $B$ physics, and will hopefully lead to the observation of new physics phenomena not accounted for in the SM.

Heavy flavour physics in the quark sector is not limited to beauty hadrons alone. The LHC is also an abundant source of charmed hadrons, which provide another interesting laboratory for BSM-physics searches. The recent experimental improvements in the measurement of mixing-related observables of $D^0$ mesons at LHCb
have raised the interest for \CP-violation measurements in this sector. Belle II and LHCb, with its upgraded detector that will be operational in the third LHC run, will probe \CP violation in charm mixing with ultimate precision. The top quark is also another excellent tool for seeking BSM physics. The unprecedented samples collected by the ATLAS and CMS experiments will enable relevant studies of \CP violation in top-quark production and decays to be carried out.

\addcontentsline{toc}{section}{References}
\setboolean{inbibliography}{true}
\bibliographystyle{LHCb}
\bibliography{KoppenburgVagnoni,exp,theory,LHCb-PAPER,LHCb-DP,LHCb-TDR}

\end{document}